\documentclass{article}
\usepackage[preprint]{neurips_2026}

\usepackage[utf8]{inputenc}
\usepackage[T1]{fontenc}
\usepackage{hyperref}
\usepackage{url}
\usepackage{booktabs}
\usepackage{amsfonts}
\usepackage{amsmath}
\usepackage{nicefrac}
\usepackage{microtype}
\usepackage{xcolor}
\usepackage{graphicx}
\usepackage{subcaption}
\usepackage{multirow}
\usepackage{makecell}
\usepackage{enumitem}
\usepackage{algorithm}
\usepackage[noend]{algpseudocode}
\usepackage{colortbl}
\usepackage{pifont}
\usepackage{tikz}
\usetikzlibrary{positioning,arrows.meta,calc,fit,shadows,shapes.geometric}
\usepackage{soul}
\usepackage{tabularx}
\usepackage{wrapfig}
\usepackage{float}

\definecolor{groundcolor}{RGB}{34,139,34}
\definecolor{citecolor}{RGB}{0,100,180}
\definecolor{rejectcolor}{RGB}{200,50,50}
\definecolor{fig1user}{RGB}{230,241,255}
\definecolor{fig1userbd}{RGB}{60,120,200}
\definecolor{fig1sys}{RGB}{249,249,249}
\definecolor{fig1sysbd}{RGB}{150,150,150}
\definecolor{fig1rej}{RGB}{255,232,232}
\definecolor{fig1rejbd}{RGB}{200,70,70}
\definecolor{fig1rev}{RGB}{235,248,235}
\definecolor{fig1revbd}{RGB}{70,150,70}
\definecolor{fig1hl}{RGB}{196,240,196}
\definecolor{fig1cap}{RGB}{255,246,218}
\definecolor{fig1capbd}{RGB}{210,170,40}
\definecolor{fig1scene}{RGB}{255,249,228}

\newcommand{\benchmark}{\textsc{Trace}}

\newcommand{\numdiag}{10{,}000}
\newcommand{\numpoi}{2{,}400}
\newcommand{\numpersona}{47}
\newcommand{\numreview}{34{,}208}
\newcommand{\numbaseline}{14}
\newcommand{\nummetric}{25}
\newcommand{\eg}{\emph{e.g.}}

\title{\benchmark{}: Tourism Recommendation with\\Accountable Citation Evidence}

\author{%
  \textbf{Zixu Zhao}$^{1}$ \quad
  \textbf{Sijin Wang}$^{2}$ \quad
  \textbf{Yu Hou}$^{3}$ \quad
  \textbf{Yuanyuan Xu}$^{1}$ \quad
  \textbf{Yufan Sheng}$^{1}$ \\
  \textbf{Xike Xie}$^{4}$ \quad
  \textbf{Wenjie Zhang}$^{1}$ \quad
  \textbf{Won-Yong Shin}$^{3}$ \quad
  \textbf{Xin Cao}$^{1}$ \\[4pt]
  \normalfont
  $^{1}$UNSW Sydney \quad
  $^{2}$University of Adelaide \quad
  $^{3}$Yonsei University \quad
  $^{4}$USTC \\[2pt]
  \texttt{\{zixu.zhao,yuanyuan.xu,yufan.sheng,wenjie.zhang,xin.cao\}@unsw.edu.au} \\
  \texttt{sijin.wang@adelaide.edu.au} \quad
  \texttt{\{houyu,wy.shin\}@yonsei.ac.kr} \quad
  \texttt{xkxie@ustc.edu.cn}
}

\begin{document}

\maketitle

\begin{abstract}
Tourism is a high-stakes setting for conversational recommender systems (CRS): a plausible-sounding suggestion can waste real money and trip time once a traveler acts on it. Existing CRS benchmarks primarily evaluate systems with a single Recall@$k$ score over entity mentions, and tourism-specific resources add spatial or knowledge-graph context, yet none of them couple multi-turn recommendation with verbatim review-span evidence and rejection recovery. This leaves an evaluation gap for tourism recommendation that is simultaneously \emph{trustworthy}, \emph{verifiable}, and \emph{adaptive}: recommend the right point of interest (POI) for multi-aspect preferences (such as cuisine, price, atmosphere, walking distance), justify each suggestion with verifiable evidence from prior visitors so the traveler can act without trial and error, and recover when the first recommendation is rejected mid-dialogue. We introduce \textbf{\benchmark{}}, where each item is \textit{a multi-turn tourism recommendation dialogue with review-span citations and explicit rejection turns}: \numdiag{} dialogues over \numpoi{} Yelp POIs and \numreview{} reviews across eight U.S.\ cities, paired with \numbaseline{} retrieval, planning, and LLM baselines, along with \nummetric{} metrics organized under \emph{Accuracy}, \emph{Grounding}, and \emph{Recovery}. Across these baselines, \benchmark{} reveals the \textbf{Three-Competency Gap}: LLM Zero-Shot leads in closed-set Recall@1 and rejection recovery but cites less densely than retrievers; non-LLM retrievers achieve surface-verbatim grounding but with low accuracy; Multi-Review Synthesis fails at recovery. The Grounding Score agrees with human citation precision (Spearman $\rho{=}{+}0.80$, $p{<}10^{-20}$), and paired $t$-tests reproduce the per-baseline ranking ($p{<}0.01$ on the dominant contrasts). \benchmark{} reframes accountable tourism recommendation as a joint target (right POI, verifiable evidence, adaptive repair) rather than a single-axis leaderboard.
\end{abstract}
\section{Introduction}
\label{sec:intro}

\begin{figure}[!t]
  \vspace{-0.5cm}
  \centering
  \includegraphics[width=\linewidth]{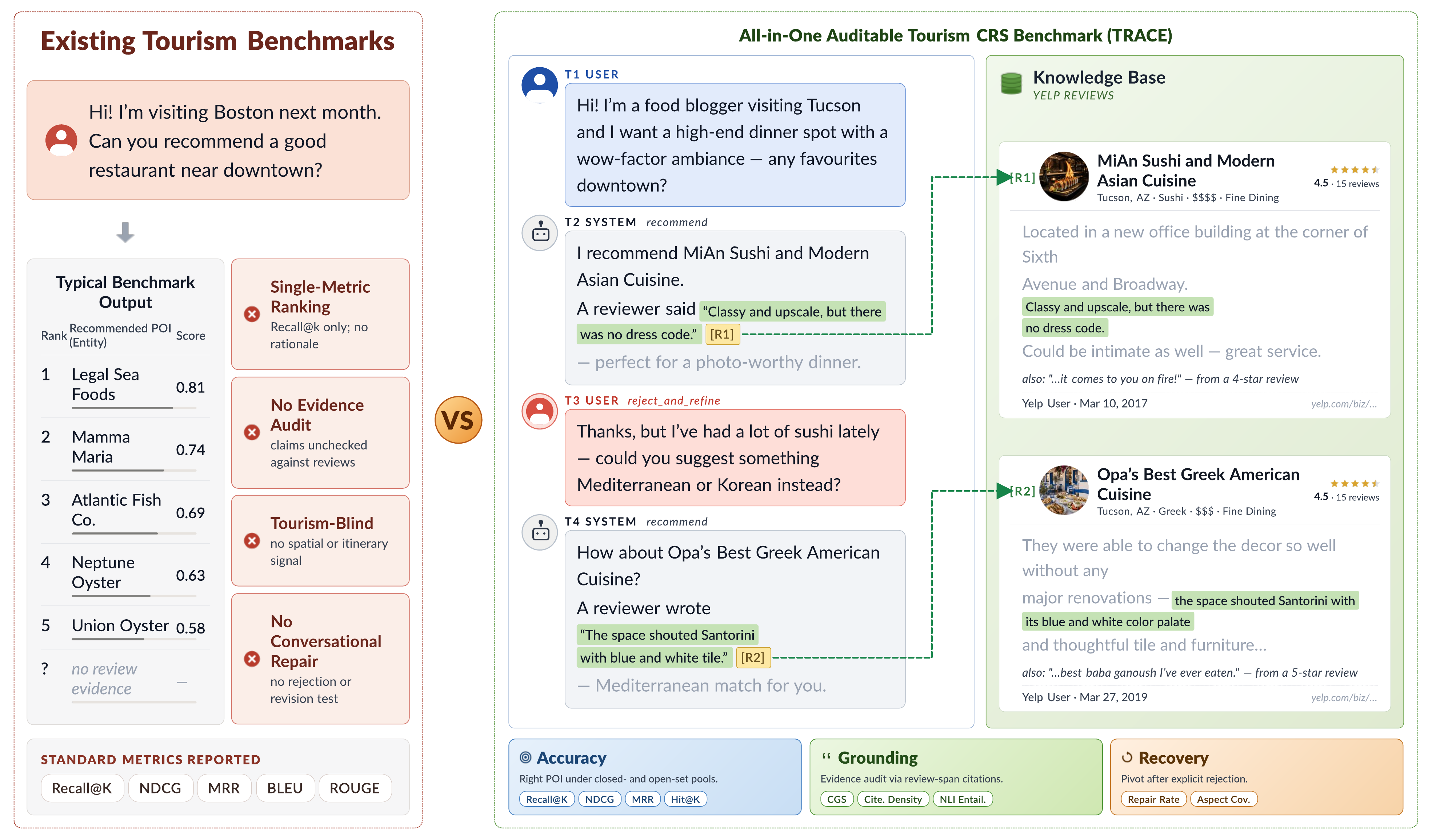}
  \caption{\textbf{Left:} four gaps in current tourism benchmarks (\emph{Single-Metric Ranking}, \emph{No Evidence Audit}, \emph{Tourism-Blind}, \emph{No Conversational Repair}). \textbf{Right:} \benchmark{} merges the three missing audits into a multi-turn dialogue with review-span citations, scoring \emph{Accuracy}, \emph{Grounding}, \emph{Recovery}.}
  \label{fig:example_dialogue}
  \vspace{-0.5cm}
\end{figure}
Tourism is a high-stakes setting for conversational recommendation: a bad restaurant suggestion in a city the traveler will not revisit costs real money and irreplaceable trip time, and a refresh cannot undo the mistake. A traveler asks three things of a conversational recommender system (CRS) at once: the \emph{right} point of interest (POI) for stated preferences (\textbf{accuracy}), a \emph{verifiable rationale} backed by prior-visitor evidence rather than fluent assertion (\textbf{grounding}), and \emph{recovery} when the first suggestion is rejected mid-dialogue (\textbf{recovery}).

\textbf{Why existing evaluations fall short.}
The left panel of Figure~\ref{fig:example_dialogue} highlights four gaps.
\emph{(i) Tourism-blind.} Travelers must balance spatial coherence, multi-aspect preferences, and itinerary feasibility, while tourism-specific resources such as DTCRSKG~\citep{fang2022dtcrskg}, TourismQA~\citep{contractor2021poi}, and RETAIL~\citep{li2025retail} add spatial or KG context without binding recommendations to verbatim review evidence.
\emph{(ii) Single-metric ranking.} Existing CRS benchmarks, including ReDial~\citep{li2018towards}, TG-ReDial~\citep{zhou2020towards}, INSPIRED~\citep{hayati2020inspired}, and DuRecDial 2.0~\citep{liu2021durecdial}, largely rank systems by Recall@$k$ over entity mentions, without assessing rationale quality.
\emph{(iii) No evidence audit.} Accuracy-only evaluation lets plausible but fabricated justifications go unchecked.
\emph{(iv) No conversational repair.} Existing benchmarks rarely test whether systems can recover after a traveler rejects the first suggestion mid-dialogue.

\textbf{How \benchmark{} addresses these gaps.} \benchmark{} replaces the single-metric ranking unit with a multi-turn tourism dialogue carrying review-span citations and explicit rejection turns: \numdiag{} dialogues over \numpoi{} Yelp POIs (800 each of restaurants, hotels, attractions in 8 U.S.\ cities) and \numreview{} reviews, paired with \numbaseline{} baselines (retrieval, planning, LLM, cross-session memory) and \nummetric{} metrics organized under \emph{Accuracy} (Recall@$k$, MRR, Hit@$k$), \emph{Grounding} (CGS, Citation Density, NLI entailment), and \emph{Recovery} (rejection-repair rate, Turns-to-First-Correct).

\textbf{The main finding.} No baseline wins all three axes. LLMs lead accuracy (closed-set Recall@1 up to 0.533) and recovery (LLM Zero-Shot \texttt{reject\_and\_refine} 0.988) but cite less densely than non-LLM retrievers (CD $\in$ [0.16, 0.39] for LLM-based, 0.001 for Multi-Review Synthesis, vs.\ [0.31, 0.44] for retrievers). Retrievers reach CGS up to 0.864 at Recall@1 $\le$0.257, but their grounding is largely template-forced verbatim text whose entailment falls to 0.110--0.193. Multi-Review Synthesis collapses on recovery (0.263), making the Three-Competency Gap visible on all three axes.

\textbf{Contributions.} TRACE introduces, to our knowledge, the first benchmark formulation that evaluates tourism CRS as a joint problem of preference-correct recommendation, review-span-verifiable evidence, and post-rejection repair.
\begin{enumerate}[leftmargin=*,nosep]
  \item \textbf{\benchmark{} dataset (\S\ref{sec:construction}).} \numdiag{} multi-turn tourism dialogues over a balanced \numpoi{}-POI Yelp KB and \numreview{} reviews, with verbatim review-span citations on every recommendation.
  \item \textbf{Three-competency evaluation (\S\ref{sec:evaluation}).} A joint accuracy / grounding / recovery protocol with three benchmark-specific metrics: CGS (closes the citation-free pass-through hole in fuzzy GS), NLI-based entailment grounding (catches span copying), and \texttt{reject\_and\_refine} correct-rate (targets repair behaviour absent from prior CRS evaluation).
  \item \textbf{Baselines and gap analysis (\S\ref{sec:three_competency_gap}).} Across \numbaseline{} baselines, no system wins all three axes; LLMs lead accuracy / recovery, retrievers lead surface grounding, MRS collapses on recovery; cross-family LLM checks (Haiku 4.5, Qwen3.5-Flash) replicate the LLM-family pattern.
  \item \textbf{Human-validated metrics (\S\ref{sec:human_eval_main}).} On 90 items with five trained annotators, GS / CGS / CD correlate positively with human citation precision (Spearman $\rho{=}{+}0.80$, $+0.47$, $+0.29$); paired $t$-tests confirm the per-baseline ordering.
\end{enumerate}

\begin{table}[!t]
  \centering
  \caption{Comparison with existing CRS benchmarks. Only \benchmark{} combines multi-turn tourism CRS, review-span evidence, spatial reasoning, and multi-aspect preference structure. Scale of recommendable inventories varies in convention across these resources (mentions vs.\ KG nodes vs.\ unique items); we therefore report \#Dial.\ for cross-comparison and defer KB scale to Table~\ref{tab:dataset_stats}.}
  \label{tab:benchmark_comparison}
  \footnotesize
  \setlength{\tabcolsep}{3pt}
  \begin{tabularx}{\linewidth}{@{}l l X c c c r@{}}
    \toprule
    \textbf{Benchmark} & \textbf{Domain} & \textbf{Grounding} & \textbf{Review} & \textbf{Spatial} & \textbf{Multi-} & \textbf{\#Dial.} \\
    & & \textbf{Type} & \textbf{Evid.} & & \textbf{aspect} & \\
    \midrule
    \multicolumn{7}{@{}l}{\emph{General-domain CRS (spatial N/A)}} \\
    ReDial~\citep{li2018towards} & Movie & None & \ding{55} & --- & \ding{55} & 10,006 \\
    TG-ReDial~\citep{zhou2020towards} & Movie & Topic & \ding{55} & --- & \ding{55} & 10,000 \\
    INSPIRED~\citep{hayati2020inspired} & Movie & Strategy & \ding{55} & --- & \ding{55} & 1,001 \\
    DuRecDial 2.0~\citep{liu2021durecdial} & Multi & KG triple & \ding{55} & --- & \ding{55} & 16,482 \\
    OpenDialKG~\citep{moon2019opendialkg} & Multi & KG path & \ding{55} & --- & \ding{55} & 15,673 \\
    \midrule
    \multicolumn{7}{@{}l}{\emph{POI / tourism resources (spatial applicable)}} \\
    DTCRSKG~\citep{fang2022dtcrskg} & Travel & KG entity & \ding{55} & \ding{55} & \ding{55} & 1{,}650\textsuperscript{$\ddagger$} \\
    TourismQA~\citep{contractor2021poi} & POI/QA & Reviews & \ding{51}\textsuperscript{$\dagger$} & \ding{55} & \ding{51} & 47{,}124\textsuperscript{$\dagger$} \\
    RETAIL~\citep{li2025retail} & Travel & Plan & \ding{55} & \ding{51} & \ding{51} & 10{,}182 \\
    \midrule
    \textbf{\benchmark{}} & \textbf{Tourism} & \textbf{Review} & \ding{51} & \ding{51} & \ding{51} & \numdiag{} \\
    \bottomrule
  \end{tabularx}
  \\[2pt]
  \footnotesize{$\dagger$ TourismQA is single-turn paragraph-level QA over tourism reviews; \#Dial. counts user questions, not multi-turn dialogues. $\ddagger$ DTCRSKG is a system; \#Dial.\ is KdConvRec (CwConvRec: 779).}
\end{table}

\section{\benchmark{}: A Three-Competency Benchmark for Accountable Tourism CRS}
\label{sec:construction}

\begin{wraptable}{r}{0.55\linewidth}
  \vspace{-\baselineskip}
  \caption{How \benchmark{} addresses the four gaps from \S\ref{sec:intro}.}
  \label{tab:gap_to_solution}
  \footnotesize
  \setlength{\tabcolsep}{3pt}
  \renewcommand{\arraystretch}{1.1}
  \resizebox{\linewidth}{!}{%
  \begin{tabular}{@{}p{0.30\linewidth}p{0.40\linewidth}p{0.30\linewidth}@{}}
    \toprule
    \textbf{Gap (\S\ref{sec:intro})} & \textbf{Mechanism (\S\ref{sec:construction})} & \textbf{Metric (\S\ref{sec:evaluation})} \\
    \midrule
    (i) Single-Metric Ranking & Multi-turn dialogue as the evaluation unit & Three axes: Accuracy, Grounding, Recovery \\
    (ii) No Evidence Audit & Verbatim review-span citation on every recommendation turn & CGS, Citation Density, NLI entailment \\
    (iii) Tourism-Blind & 8 cities $\times$ 3 POI types; spatial + multi-aspect personas & Spatial Coherence, Preference Coverage \\
    (iv) No Conversational Repair & \texttt{reject\_and\_refine} turns mid-dialogue & Rejection-repair rate, Turns-to-First-Correct \\
    \bottomrule
  \end{tabular}}
  \vspace{-\baselineskip}
\end{wraptable}

\textbf{The three competencies.}\label{sec:competencies} What should a tourism CRS have to prove before we trust a recommendation? We organize \benchmark{} around three capabilities that must hold together:
\begin{itemize}[leftmargin=*,nosep]
  \item \textbf{C1 Accuracy:} recommend the POI that best matches the user's stated preferences.
  \item \textbf{C2 Grounding:} back recommendation claims with review-span evidence from a referenced review.
  \item \textbf{C3 Recovery:} pivot to a correct recommendation after explicit user rejection.
\end{itemize}
Together these three axes define whether a CRS can be audited end-to-end: can it get the answer right, justify it, and fix itself when wrong? Table~\ref{tab:gap_to_solution} lays out the explicit mapping from each Introduction gap (Section~\ref{sec:intro}) to its construction mechanism and evaluation metric. Section~\ref{sec:evaluation} gives the metric definitions; full formulas and sensitivity checks are in Appendix~\ref{app:metrics_full}.

\subsection{Dataset Construction}
\label{sec:construction_detail}

\benchmark{} is constructed through a five-stage pipeline: scenario construction, candidate POI selection, dialogue generation, quality validation, and difficulty annotation (pipeline diagram in Appendix~\ref{app:pipeline_full}; data-collection and licensing details in Appendix~\ref{app:datasheet}).

\paragraph{Knowledge base.}
The Yelp-based knowledge base is balanced across 8 U.S.\ cities and 3 tourism POI types (restaurants, hotels, attractions). It combines real reviews with structured attributes needed for recommendation, citation, and spatial evaluation; Table~\ref{tab:dataset_stats} gives the counts and dialogue-level statistics.

\begin{table}[t]
  \caption{\benchmark{} benchmark statistics: balanced Yelp tourism KB, \numdiag{} multi-turn dialogues, 29.4\% rejection turns, and four generation-time difficulty tiers.}
  \label{tab:dataset_stats}
  \centering
  \footnotesize
  \setlength{\tabcolsep}{4pt}
  \begin{tabularx}{\linewidth}{ll@{\hspace{2em}}ll}
    \toprule
    \multicolumn{2}{l}{\textbf{Knowledge base (Yelp)}} & \multicolumn{2}{l}{\textbf{Dialogues}} \\
    \midrule
    POIs & \numpoi{} (800 ea.\ restaurant/hotel/attr.) & Total dialogues & \numdiag{} \\
    Reviews & \numreview{} (avg.\ 14.3/POI) & Avg.\ turns / dialogue & 10.5 ($\pm$1.9) \\
    Avg.\ stars & 3.8 & Avg.\ POIs rec.\ / dialogue & 3.3 \\
    Unique categories & 485 & Persona archetypes & 47 \\
    Cities & 8 (U.S., Yelp Academic) & Candidates / dialogue & 8 \\
    Price range & \$--\$\$\$\$ (restaurants, hotels) & Rejection rate & 29.4\% \\
    \midrule
    \multicolumn{2}{l}{\textbf{Difficulty (generation-time)}} & E / M / H / Expert & 2{,}481 / 3{,}476 / 3{,}063 / 980 \\
    \bottomrule
  \end{tabularx}
\end{table}

\subsection{Dialogue Generation and Quality}
\label{sec:pipeline}

\textbf{Generator.} Stages~1--3 sample one of 47 persona archetypes, select 8 candidate POIs by stratified rating/category sampling, and generate the dialogue with DSPy~\citep{khattab2023dspy} and \texttt{gpt-5.4-mini}. The prompt requires strict user--system alternation, at least 3 distinct recommendations, at least 1 \texttt{compare} turn, tier-specific \texttt{reject\_and\_refine} turns, and verbatim review citations via \texttt{[Rn]} labels. Stages~4--5 validate turn structure, action coverage, POI/review references, and per-quote fuzzy match ($\geq$0.80 partial ratio), then assign Easy/Medium/Hard/Expert difficulty labels. Full prompts, action vocabulary, retry policy, and the exploratory E3 score are in Appendix~\ref{app:difficulty}.

\textbf{Multi-reference gold and open-set candidates.} Each dialogue ships with a single-reference gold POI set plus a \emph{multi-reference} gold set: for every recommendation turn the generator lists any other candidate in the dialogue's 8-POI pool that would also have satisfied the user's stated preferences, together with an open-set candidate list (the same-city/type POIs beyond the closed 8). This lets evaluators run the same dialogues under closed (8), expanded (16, 32), or open-set ($\sim$50--100 same-city/type POIs) candidate pools without regenerating.

\textbf{Data quality checks.} For semantic quality we run an ALCE-style human evaluation (details Appendix~\ref{app:human_eval}) on 90 items (30 dialogues $\times$ 3 baselines: TF-IDF, RAG-Citation, LLM Zero-Shot) with five trained annotators. Krippendorff's $\alpha$ is moderate on Likert scales ($\alpha_{\text{info}}{=}0.58$, $\alpha_{\text{nat}}{=}0.42$) and lower on nominal citation / error labels ($\alpha_{\text{cite}}{=}0.32$, $\alpha_{\text{err}}{=}0.31$); these all-five-annotator values are reported without post-hoc filtering, and Appendix~\ref{app:human_eval} includes a leave-one-out annotator diagnostic. The qualitative ordering among the three audited baselines is stable under majority vote (\S\ref{sec:human_eval_main}).

\textbf{Release.} Our authored content (dialogue structure, non-quote turn text, action labels, grounding annotations, \texttt{(business\_id, R\#)} citation pointers, the \numbaseline{} baselines, and evaluation toolkit) is released under \textbf{CC BY 4.0}. Verbatim Yelp review spans are \emph{not} redistributed: they appear as \texttt{[Q:N]} placeholders with positional metadata (review ID + char offsets) and are hydrated locally from a user-obtained Yelp Open Dataset\footnote{\url{https://business.yelp.com/data/resources/open-dataset/}} bundle under Yelp's academic-use terms (non-commercial, non-sublicensable, revocable; \S3, \S4.A, \S4.H, \S10 of the Feb 2021 agreement). POI metadata (name, address, lat/lon, categories, stars, price) is included under the \S4.E academic-publication carve-out; \texttt{business\_id}/\texttt{review\_id} are bare factual identifiers, and the full \numreview{} review corpus is not redistributed. Official splits are city$\times$POI-type stratified 70/15/15 (\texttt{seed=42}); field inventory and licensing in Appendix~\ref{app:datasheet}.\footnote{Anonymous repository: \url{https://anonymous.4open.science/r/TRACE-benchmark}.}

\begin{table}[h]
  \vspace{-0.3\baselineskip}
  \caption{\textbf{Reusability checklist}. ``Recurrent status'' summarizes the prior resources cited in Table~\ref{tab:benchmark_comparison}.}
  \label{tab:reusability}
  \footnotesize
  \setlength{\tabcolsep}{4pt}
  \renewcommand{\arraystretch}{1.05}
  \begin{tabularx}{\linewidth}{@{}p{1.7cm} X X@{}}
    \toprule
    \textbf{Factor} & \textbf{Recurrent status} & \textbf{\benchmark{}} \\
    \midrule
    Access & gated forms, broken URLs, no install path; require re-scrape & one-command install + anonymous mirror; HF Hub release with DOI; \texttt{reconstruct\_kb.py} hydrates Yelp spans locally \\
    Metadata & README only; quote$\to$source link implicit & datasheet \citep{gebru2021datasheets} + JSON schema; per-quote pointer \texttt{(business\_id, R\#, offsets)} \\
    Versioning & undated, unversioned dumps; silent KB drift & KB / dialogue versions tagged with SHA256; POI IDs stable across KB versions \\
    Eval.\ repro. & paper-only metrics; baselines absent; seeds undocumented & \numbaseline{} baselines + \nummetric{} metrics shipped + CLI; \texttt{seed=42}; open-set, multi-ref, cross-judge, sensitivity \\
    Legal & MIT / Apache-2.0 claims that conflict with Yelp/TripAdvisor non-sublicensable ToU & CC BY 4.0 on synthetic content; verbatim Yelp text \emph{not} redistributed (\texttt{[Q:N]} placeholders); POI metadata under ToU \S4.E \\
    \bottomrule
  \end{tabularx}
  \vspace{-0.6\baselineskip}
\end{table}

\paragraph{Two difficulty knobs.}\label{sec:difficulty_knobs} \benchmark{} exposes two orthogonal axes. \textbf{(K1) Generation tier} (Easy/Medium/Hard/Expert, 24.8/34.8/30.6/9.8\,\%; Table~\ref{tab:dataset_stats}) sets intrinsic dialogue complexity. \textbf{(K2) Pool size} scales retrieval difficulty on fixed dialogues from 8 (closed) to 16, 32, $\sim$50--100 (open). This lets the same dialogue test dialogue complexity and retrieval difficulty independently; intermediate sweeps are in Appendix~\ref{app:pool_ablation}.

\section{Evaluation Framework}
\label{sec:evaluation}

\benchmark{} is designed so no single metric family can stand in for system quality. We evaluate \nummetric{} metrics in seven families (full catalog in Appendix~\ref{app:catalogs}, Table~\ref{tab:metrics}): text quality, recommendation quality, grounding, practical value, dialogue recovery, LLM-judge ratings, and human evaluation. Standard metrics follow conventional definitions; novel grounding and practical-value metrics are summarized below and specified in Appendix~\ref{app:metrics_full}.

\paragraph{Novel grounding metrics.}
CGS multiplies quote fidelity, citation density, and provenance coverage so citation-free responses cannot receive grounding credit: $\text{GS}\times\min(1,\text{CD}/0.05)\times(0.5+0.5\times\text{PC})$. GS fuzzy-matches quoted substrings to cited reviews, CD counts verbatim spans from cited reviews, and PC checks whether aspect mentions have nearby citations. Entailment Grounding uses DeBERTa-v3-base NLI (threshold 0.5) for paraphrased claims; component orthogonality and CGS sensitivity are in Appendix~\ref{app:cgs_sensitivity}.

\paragraph{Novel practical-value metrics.}
\textbf{Spatial Coherence} is the fraction of recommended POI pairs within 2\,km (about a 25\,min walk), with 1--5\,km sensitivity in Appendix~\ref{app:sensitivity}. \textbf{Price Alignment} maps budget keywords to price ceilings (budget/cheap/affordable $\to$ \$--\$\$; moderate $\to$ \$--\$\$\$; luxury/upscale $\to$ \$--\$\$\$\$), reaches 96\% extraction recall on 100 manually checked dialogues, and excludes the 38\% without explicit price. Full aggregation details are in Appendix~\ref{app:metrics_full}.

\section{Baseline Systems}
\label{sec:baselines}

The baseline set isolates retrieval, state tracking, citation constraints, aspect synthesis, spatial planning, and cross-session memory. We test \numbaseline{} baselines in four groups (full catalog in Appendix~\ref{app:catalogs}, Table~\ref{tab:baselines}): \textbf{Non-LLM Retrieval} (Popularity, TF-IDF, Aspect, Dense), \textbf{LLM-Direct} (LLM Zero-Shot), \textbf{LLM+Retrieve} (DST, RAG-Citation, Multi-Review Synthesis), and \textbf{LLM+Plan/Reflect} (Itinerary-style structure and Memory-Augmented cross-session preference memory). The 13 session-isolated baselines are used in Tables~\ref{tab:main_results}--\ref{tab:stress}; Memory-Aug.\ is reported separately because it reads persona-keyed preferences from prior dialogues and is therefore not session-isolated.

\paragraph{Key baseline mechanisms.}
Non-LLM baselines cover popularity, sparse TF-IDF, dense retrieval, spatial blending, hybrid fusion, itinerary routing, attribute matching, and persona-aware reranking. LLM baselines use \texttt{gpt-5.4-mini} as the primary backbone: Zero-Shot, DST preference-state filtering, RAG-Citation with a quote constraint, Multi-Review Synthesis by aspect consensus, and Memory-Aug.\ with a 168h half-life cross-session memory over \numpersona{} persona types ($\sim$213 dialogues each). LLM Zero-Shot is additionally run with two cross-family backbones (Anthropic Haiku~4.5, Qwen3.5-Flash) on a 1{,}000-dialogue subset to verify family-level effects (\S\ref{sec:experiments}; Table~\ref{tab:main_results} bottom rows). Memory-Aug.\ is not session-isolated; all others are. Details, hyperparameters, and executable configs are in Appendix~\ref{app:baselines}.

\section{Experimental Setup}
\label{sec:experiments}
\label{sec:setup}

\textbf{Models.} The five LLM baselines (Zero-Shot, DST, RAG-Citation, Multi-Review Synthesis, Memory-Aug.) use \texttt{gpt-5.4-mini} via DSPy as the primary backbone on the full \numdiag{}-dialogue corpus; dense retrieval uses \texttt{all-MiniLM-L6-v2}. To check that the LLM-family pattern is not a single-model artifact, LLM Zero-Shot is additionally run with two cross-family backbones, Anthropic Haiku~4.5 and Qwen3.5-Flash, on a stratified 1{,}000-dialogue subset (Table~\ref{tab:main_results} bottom rows; \S\ref{sec:c1_accuracy}). We do \emph{not} run a fully controlled $\numbaseline\times M$ multi-LLM sweep at this stage: with five LLM baselines $\times$ \numdiag{} dialogues $\times$ multiple recommend / compare / reject-and-refine turns per dialogue, the per-backbone API cost exceeds a single review cycle, and a fair comparison would also require per-backbone prompt re-tuning. The released evaluation toolkit lets users swap the backbone via a single DSPy config flag, so a controlled multi-LLM leaderboard is the natural next deliverable for the camera-ready release. All dataset / evaluation sampling uses \texttt{random.seed=42}; DSPy/litellm decoding uses one response, default retries, \texttt{max\_tokens=600} for non-reasoning models, and \texttt{temperature=1.0}, \texttt{max\_tokens=16{,}000} for \texttt{gpt-5.4-mini}.

We evaluate the full \numdiag{} training-free corpus in closed set (the generation-time 8-POI pool) and report open set on a 1{,}000-dialogue subset over all same-city/type POIs in the \numpoi{}-POI KB ($\sim$50--100 candidates). Because each baseline exposes a different candidate interface, open-set results are system-level rather than model-only comparisons: LLM Zero-Shot/RAG-Citation use TF-IDF top-16, DST/MRS rerank the full pool internally, and non-LLM baselines see the full pool.

Auxiliary protocols include entailment grounding (non-LLM full corpus, LLM 1{,}000-dialogue subset, $n=5{,}245$ turns/baseline), multi-reference, 90-item human evaluation, and 200-dialogue cross-judge; the official 1{,}496-dialogue test split is reserved for future trained baselines. In headline tables, \textbf{bold} and \underline{underline} mark best and second-best values; lower-is-better columns follow the arrow in the header.

\section{The Three-Competency Gap}
\label{sec:three_competency_gap}

We test \benchmark{} along the three competencies introduced in \S\ref{sec:competencies}: accuracy (\S\ref{sec:c1_accuracy}), grounding (\S\ref{sec:c2_grounding}), and recovery (\S\ref{sec:c3_recovery}). Across \numbaseline{} baselines, no single system wins all three: Table~\ref{tab:main_results} reports the closed-set headline across nine metrics; Table~\ref{tab:stress} reports per-axis stressors (open-set R@1, fuzzy and entailment grounding, rejection recovery). LLMs lead closed-set accuracy and rejection recovery, non-LLM retrievers dominate surface-verbatim grounding at low accuracy, and Multi-Review Synthesis collapses on recovery. Figure~\ref{fig:tradeoff} synthesizes the gap on the open-set projection.

\begin{table}[t]
  \vspace{-0.5cm}
  \caption{\textbf{Main results} on the full \numdiag{}-dialogue corpus (closed-set, 8 candidates, session-isolated). LLM Zero-Shot leads on accuracy, text quality, and recovery but trails non-LLM retrievers on grounding and citation density. \textbf{Bold} = best, \underline{underline} = runner-up among the 13 baselines; cross-family rows ($\dagger$, 1{,}000-dialogue subset) confirm the pattern is not model-specific.}
  \label{tab:main_results}
  \centering
  \footnotesize
  \setlength{\tabcolsep}{3pt}
  \begin{tabular*}{\linewidth}{@{\extracolsep{\fill}}lccccccccc}
    \toprule
    & \multicolumn{3}{c}{\textbf{Accuracy}} & \multicolumn{2}{c}{\textbf{Text Quality}} & \multicolumn{2}{c}{\textbf{Grounding}} & \multicolumn{2}{c}{\textbf{Recovery}} \\
    \cmidrule(lr){2-4} \cmidrule(lr){5-6} \cmidrule(lr){7-8} \cmidrule(lr){9-10}
    \textbf{Baseline} & R@1$\uparrow$ & R@3$\uparrow$ & MRR$\uparrow$ & BLEU$\uparrow$ & ROU.$\uparrow$ & CGS$\uparrow$ & Cite.D.$\uparrow$ & TSR$\uparrow$ & T2R$\downarrow$ \\
    \midrule
    \multicolumn{10}{l}{\emph{Non-LLM retrievers / planners}} \\
    Popularity      & 0.140 & 0.189 & 0.164 & 0.078 & 0.105 & \underline{0.802} & 0.411 & 0.734 & 5.57 \\
    TF-IDF          & 0.257 & 0.321 & 0.289 & 0.095 & 0.119 & 0.776 & 0.377 & 0.948 & 4.02 \\
    Aspect          & 0.179 & 0.232 & 0.206 & 0.088 & 0.110 & 0.750 & 0.391 & 0.814 & 4.96 \\
    Dense           & 0.243 & 0.305 & 0.274 & 0.082 & 0.110 & \textbf{0.864} & 0.393 & 0.915 & 4.17 \\
    Spatial         & 0.257 & 0.320 & 0.288 & 0.077 & 0.111 & 0.727 & \textbf{0.437} & 0.949 & 4.02 \\
    Hybrid-RRF      & 0.254 & 0.317 & 0.286 & 0.076 & 0.103 & 0.775 & \underline{0.412} & 0.938 & 4.01 \\
    Itinerary       & 0.257 & 0.321 & 0.289 & 0.099 & 0.108 & 0.684 & 0.310 & 0.948 & 4.02 \\
    Knowledge-Enh.  & 0.206 & 0.262 & 0.234 & 0.107 & 0.108 & 0.708 & 0.375 & 0.842 & 4.70 \\
    Persona-Ground. & 0.191 & 0.245 & 0.218 & 0.093 & 0.115 & 0.706 & 0.354 & 0.828 & 4.83 \\
    \midrule
    \multicolumn{10}{l}{\emph{LLM baselines}} \\
    LLM Zero-Shot   & \textbf{0.533} & \textbf{0.754} & \textbf{0.638} & \textbf{0.130} & \textbf{0.198} & 0.636 & 0.162 & \textbf{1.000} & \textbf{3.00} \\
    DST             & \underline{0.442} & \underline{0.579} & \underline{0.508} & \underline{0.120} & \underline{0.177} & 0.634 & 0.194 & \underline{0.959} & \underline{3.38} \\
    RAG-Citation    & 0.381 & 0.436 & 0.409 & 0.106 & 0.159 & 0.658 & 0.388 & 0.949 & 3.69 \\
    Multi-Rev.\ Synth. & 0.281 & 0.297 & 0.289 & 0.083 & 0.104 & 0.008 & 0.001 & 0.521 & 4.55 \\
    \cmidrule(lr){1-10}
    \multicolumn{10}{l}{\emph{LLM Zero-Shot cross-family validation $^\dagger$}} \\
    \quad Claude Haiku 4.5      & 0.485 & 0.607 & 0.545 & 0.102 & 0.175 & 0.541 & 0.159 & 0.957 & 3.37 \\
    \quad Qwen3.5-Flash  & 0.477 & 0.622 & 0.548 & 0.099 & 0.162 & 0.505 & 0.100 & 0.986 & 3.15 \\
    \bottomrule
  \end{tabular*}

  \vspace{2pt}
  \footnotesize $^\dagger$Cross-family rows: 1{,}000-dialogue subset, excluded from the ranks above (which span the full \numdiag{}-corpus).
  \vspace{-0.5cm}
\end{table}

\subsection{Competency 1, Accuracy: LLMs Lead Closed-Set, Lose Most of It Open-Set}
\label{sec:c1_accuracy}

\paragraph{Question.} Given the user's preferences expressed across multiple dialogue turns, does the system pick the right POI? We measure Recall@1, Recall@3, and MRR against the single-reference gold POI for each recommendation turn.

\textbf{Finding 1a: LLMs lead closed-set accuracy.} LLM Zero-Shot leads at Recall@1/MRR = 0.533/0.638; DST trails at 0.442/0.508. Non-LLM retrievers cluster near TF-IDF (R@1 = 0.257), with sparse, spatial, RRF, and itinerary within 0.003, while popularity/aspect stay below 0.18. Spatial filtering, attribute matching, and persona reweighting beat popularity but leave a $>$2$\times$ LLM-vs-retriever ceiling gap (Table~\ref{tab:stress}).

\begin{wraptable}{r}{0.55\linewidth}
  \vspace{-\baselineskip}
  \caption{\textbf{Per-axis stress tests.} Open-set R@1 on 1{,}000-dialogue subset; GS/PC/NLI on full corpus. \textbf{Bold} = best, \underline{underline} = runner-up. $^\ast$MRS PC defaults to 1.0; Knowledge-Enh./Persona-Ground.\ NLI omitted (Appendix~\ref{app:metrics_full}).}
  \label{tab:stress}
  \centering
  \resizebox{\linewidth}{!}{%
  \begin{tabular}{lccccc}
    \toprule
    \textbf{Baseline} & \textbf{R@1 Opn}$\uparrow$ & \textbf{GS}$\uparrow$ & \textbf{PC}$\uparrow$ & \textbf{NLI}$\uparrow$ & \textbf{Rej.\ Rec.}$\uparrow$ \\
    \midrule
    \multicolumn{6}{l}{\emph{Non-LLM retrievers / planners}} \\
    Popularity      & 0.012 & 0.998 & 0.608 & 0.153 & 0.626 \\
    TF-IDF          & 0.099 & 0.998 & 0.555 & \textbf{0.193} & 0.891 \\
    Aspect          & 0.017 & 0.998 & 0.503 & \underline{0.182} & 0.715 \\
    Dense           & 0.033 & 0.997 & \textbf{0.733} & 0.168 & 0.821 \\
    Spatial         & 0.099 & 0.997 & 0.458 & 0.143 & 0.893 \\
    Hybrid-RRF      & 0.078 & 0.997 & 0.555 & 0.171 & 0.854 \\
    Itinerary       & 0.099 & 0.997 & 0.373 & 0.110 & 0.891 \\
    Knowledge-Enh.  & 0.052 & \textbf{0.999} & 0.417 & --- & 0.752 \\
    Persona-Ground. & 0.018 & \textbf{0.999} & 0.413 & --- & 0.737 \\
    \midrule
    \multicolumn{6}{l}{\emph{LLM baselines}} \\
    LLM Zero-Shot   & \textbf{0.132} & 0.981 & 0.297 & 0.038 & \textbf{0.988} \\
    DST             & \underline{0.125} & 0.982 & 0.291 & 0.044 & \underline{0.895} \\
    RAG-Citation    & 0.093 & 0.994 & 0.324 & 0.033 & 0.863 \\
    Multi-Rev.\ Synth. & 0.100 & 0.996 & \underline{0.980}$^\ast$ & 0.050 & 0.263 \\
    \bottomrule
  \end{tabular}%
  }
  \vspace{-0.8\baselineskip}
\end{wraptable}

\textbf{Finding 1b: Open-set retrieval shrinks every baseline; the LLM lead survives.} Moving to $\sim$50--100 same-city/type POIs breaks every system (Table~\ref{tab:stress}, R@1 Opn). LLM Zero-Shot drops $0.533\to0.132$ ($-$0.401), DST $0.442\to0.125$ ($-$0.317), RAG-Citation $0.381\to0.093$ ($-$0.288), and TF-IDF/Spatial/Itinerary $0.257\to0.099$ ($-$0.158; $-$61\,\% vs.\ LLM Zero-Shot's $-$75\,\%). LLM Zero-Shot still leads open-set R@1 by 33\,\% over the best non-LLM; Dense drops hardest among non-LLM ($-$0.210).

\textbf{Finding 1c: Single-reference gold materially undercounts accuracy.} Single-reference Recall is a lower bound, not an absolute accuracy measure. Multi-Reference Recall@1 lifts all 13 session-isolated baselines by +9 to +14\,pp on average and +15 to +23\,pp on turns with $\geq$1 alternative. MRS rises 0.281$\to$0.405 (+12.5\,pp), LLM Zero-Shot 0.532$\to$0.666 (+13.4\,pp), and RAG-Citation 0.381$\to$0.519 (+13.7\,pp); within-family rankings stay fixed.

\textbf{Finding 1d: A spatial-reasoning ceiling.} Injecting the template Itinerary's 500\,m proximity graph into LLM Zero-Shot fails to raise accuracy (Itinerary-LLM closed R@1=0.538 vs.\ 0.533; open 0.083 vs.\ matched 0.088); the template's 0.640 closed-set Spatial Coherence comes from an explicit nearest-neighbor TSP step, not spatial-aware ranking, suggesting a representational-vs.-algorithmic gap (Appendix~\ref{app:pool_ablation}).

\subsection{Competency 2, Grounding: Retrievers Lead on Surface Evidence, LLMs Pay an Accuracy Tax}
\label{sec:c2_grounding}

\paragraph{Question.} When the system recommends a POI, can each recommendation claim be traced to source reviews? We report CGS plus Grounding Score (GS, verbatim fuzzy match), Citation Density (CD), Provenance Coverage (PC), and NLI-based entailment grounding (full definitions Appendix~\ref{app:metrics_full}).

\begin{wrapfigure}{r}{0.50\linewidth}
  \vspace{-\baselineskip}
  \centering
  \includegraphics[width=\linewidth]{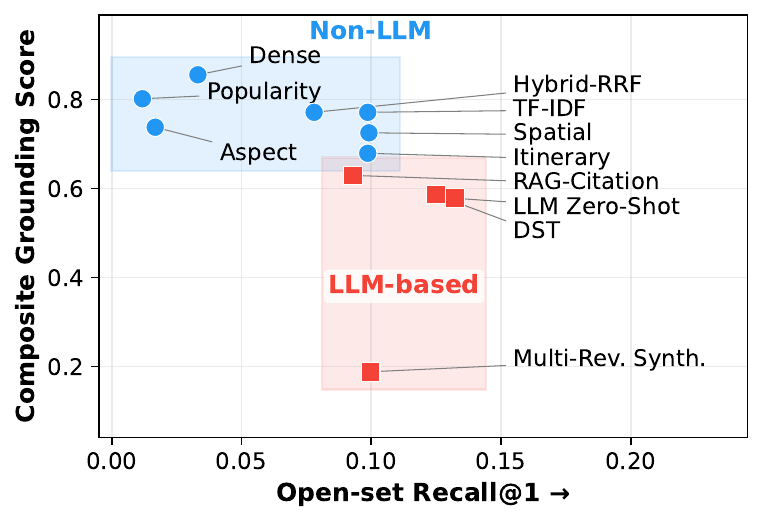}
  \caption{\textbf{Three-Competency Gap (open-set).} \textcolor[HTML]{2196F3}{Non-LLM retrievers} (grounding-led, CGS$>$0.65) and \textcolor[HTML]{F44336}{LLM-based} systems (accuracy-led) occupy disjoint zones; no baseline reaches both.}
  \label{fig:tradeoff}
  \vspace{-\baselineskip}
\end{wrapfigure}

\textbf{Finding 2a: Non-LLM retrievers own \emph{surface-verbatim} grounding.} All 9 non-LLM baselines exceed CGS=0.68, led by Dense 0.864, Popularity 0.802, and TF-IDF 0.776. RAG-Citation is the strongest LLM at 0.658, still below Persona-Grounded at 0.706; LLM Zero-Shot/DST are 0.636/0.634, and MRS collapses to 0.008 because CD=0.001. Citation density separates families: [0.31,0.44] for non-LLM, [0.16,0.39] for citing LLMs, near zero for MRS; the $\geq$10-token RAG-Citation constraint keeps CD high, while Gemini 2.5 Flash shifts LLM Zero-Shot CD by only $\approx$0.01.

\textbf{Component caveat.} We report raw GS/CD/PC alongside CGS because the composite uses hand-chosen gates (the $\text{CD}/0.05$ piece) and a PC term that can default to 1.0 on no-aspect responses. A single-number grounding leaderboard would hide which citation behavior each baseline wins or loses on.

\textbf{Finding 2b: High grounding under fuzzy match is partially vacuous.} High fuzzy grounding can mean the system copied a span, not that its full recommendation is semantically supported. Non-LLM fuzzy GS is 0.997--0.999, but entailment grounding falls to 0.110--0.193; LLM entailment is 0.029--0.050 on the 1{,}000-dialogue subset ($n=5{,}245$, \texttt{gpt-5.4-mini}), with Knowledge-Enhanced/Persona-Grounded omitted because they share TF-IDF-level fuzzy GS. The CD gate partly corrects this: CD$<$0.05 is zeroed, pulling the non-LLM advantage toward 0.13--0.19 of CGS margin rather than the raw 4$\times$ fuzzy-GS gap. Figure~\ref{fig:tradeoff} shows the open-set projection: Dense reaches CGS 0.855 at low R@1, LLM Zero-Shot leads R@1 at 0.132 (vs.\ TF-IDF/Spatial/Itinerary 0.099), and MRS sits bottom-left (open CGS 0.188; closed CGS 0.008). As scoped in \S\ref{sec:setup}, open-set axes are system-level.

\subsection{Competency 3, Recovery: Instruction-Following Helps, Aspect-Synthesis Collapses}
\label{sec:c3_recovery}

\paragraph{Question.} When the user explicitly rejects a recommendation with a new constraint, does the system's next recommendation satisfy that constraint? We measure dialogue-level Task Success Rate (TSR, whether any later turn contains a correct recommendation), Turns-to-First-Correct (T2R), and the \emph{Rejection Recovery} rate (fraction of \texttt{reject\_and\_refine} turns followed by a correct recommendation).

\textbf{Finding 3a: LLM Zero-Shot dominates recovery; aspect synthesis collapses.} LLM Zero-Shot recovers from 98.8\,\% of \texttt{reject\_and\_refine} turns, with TSR=1.000 and T2R=3.00; DST follows at 0.895, TF-IDF/Itinerary/Spatial cluster near 0.89, and RAG-Citation reaches 0.863. MRS fails at 0.263 because aspect-consensus frequencies barely change after one rejection.

\textbf{Finding 3b: Recovery appears family-sensitive in a uncontrolled rerun.} A Gemini~2.5~Flash rerun on a smaller earlier 1{,}200-POI KB reverses the headline order: template retrievers lead at $\approx$0.978, while LLM Zero-Shot/DST trail at 0.812/0.702, versus \texttt{gpt-5.4-mini} on the full \numpoi{}-POI KB. Because LLM and KB both changed, this is suggestive only; non-LLM stability (TF-IDF $0.978\to0.891$, Spatial $0.978\to0.893$, Popularity 0.626) points to but does not isolate an LLM-instruction-following effect, so recovery should be reported across LLM families when architectural claims depend on it.

\section{Human Evaluation and Discussion}
\label{sec:discussion}

\subsection{Human Evaluation}
\label{sec:human_eval_main}

\textbf{Calibration role.} The human evaluation checks whether the automated axes correspond to visible quality tradeoffs. In an ALCE-style~\citep{gao2023enabling} study, five trained annotators rated 90 items (30 dialogues $\times$ TF-IDF / RAG-Citation / LLM Zero-Shot) and wrote blind recommendations on 20 dialogues; agreement is modest ($\alpha\in[0.31,0.58]$) and all five annotators are kept in the headline values. Likert means follow automated quality (LLM Zero-Shot $>$ RAG-Citation $>$ TF-IDF on informativeness and naturalness) but citation precision inverts the ordering (RAG-Citation / TF-IDF $>$ LLM Zero-Shot), confirming that fluent or quote-heavy outputs alone are not evidence of genuine support. Full Likert / citation tables, Task 2 expert-recommendation statistics, leave-one-out IAA, and an earlier pilot are in Appendix~\ref{app:human_eval}.

\textbf{Metric--human validation.} Pooled Spearman correlation against human citation precision validates the grounding metrics' direction: GS is the strongest single proxy ($\rho{=}{+}0.80$), CGS damps it via the CD/PC gates ($\rho{=}{+}0.47$), and CD alone is weakest ($\rho{=}{+}0.29$, all $p{<}0.01$). CD and GS correlate negatively with human informativeness ($-0.60$ and $-0.25$), reflecting the LLM-vs-retriever divide at the item level. Per-dialogue paired $t$-tests confirm the headline ordering on informativeness ($p{<}0.005$ across all three pairs) and reverse it on CGS / CD, consistent with the Three-Competency Gap. Full correlation matrix and pairwise $t$-statistics are in Appendix~\ref{app:human_eval_corr}; numbers are reproducible with \texttt{scripts/compute\_human\_metric\_correlation.py}.

\subsection{Discussion: the Three-Competency Gap}

The Three-Competency Gap is not one bad baseline; it is a family-level split. LLM Zero-Shot leads closed-set accuracy/recovery (0.533/0.988) but trails all 9 non-LLM baselines on CGS (0.636); RAG-Citation raises LLM grounding to 0.658 at $-$0.15 R@1; MRS collapses on grounding/recovery (0.008/0.263). Accuracy-only leaderboards hide citation failures, CGS needs component/entailment checks, and recovery is model-family sensitive. The benchmark implication is direct: systems must be compared jointly across recommendation quality, evidence support, and repair.

\section{Related Work}
\label{sec:related}

\textbf{Conversational recommendation.} The missing CRS benchmark unit is review-level evidence and post-rejection repair. Existing benchmarks span entity-level dialogue (ReDial~\citep{li2018towards}, INSPIRED~\citep{hayati2020inspired}), topic threading (TG-ReDial~\citep{zhou2020towards}), goal-driven multi-domain recommendation (DuRecDial / DuRecDial~2.0~\citep{liu2020towards,liu2021durecdial}), and KG-grounded paths (OpenDialKG~\citep{moon2019opendialkg}); recent systems add knowledge fusion, persona/contextual adaptation, reasoning, multimodal interaction, language-modelled item interaction, and synthetic CRS dialogue~\citep{wang2022unicrs,ren2024kecr,yang2022mese,lan2023trea,kim2024pearl,kostric2025tailor,kemper2024rarec,anwaar2026reasoning,bismay2025reasoningrec,guo2025vogue,banerjee2025collabrec,rabiah2025bridging,yang2024ilm,banerjee2025synthtrips}. CRS surveys~\citep{jannach2021survey,deldjoo2024genrecsys,mahmud2026recommendation} flag citation faithfulness as unresolved, but existing grounding stops at entities, topics, KG triples, or strategy labels, not verbatim snippets that a recommendation can be audited against.

\textbf{Citation, grounding, and faithfulness.} Citation work gives \benchmark{} its audit vocabulary, but misses multi-turn preference elicitation and recommendation ranking. ALCE~\citep{gao2023enabling} introduces citation precision/recall for long-form QA; ASQA~\citep{stelmakh2022asqa} targets ambiguous factoids; FaithDial~\citep{dziri2022faithdial} scores faithful information-seeking dialogue; FACTS~\citep{jacovi2025facts}, Trace-evidence~\citep{fang2024traceevidence}, RAGLens~\citep{xiong2025raglens}, HalluGuard~\citep{zeng2026halluguard}, and ReXHa~\citep{sun2025rexha} add hallucination-detection and evidence-auditing refinements. Cite-before-speak~\citep{zeng2025cite} orders citation before generation; FinRATE~\citep{jiang2026finrate} ports the protocol to financial QA. RAG~\citep{lewis2020retrieval} is the shared primitive, and \citet{huang2024citation} survey citation behaviour in LLMs. These resources do not test whether cited evidence supports a recommendation in a later-turn CRS decision.

\textbf{Tourism, POI recommendation, and spatial reasoning.} Tourism resources miss the joint setting of multi-turn CRS, review evidence, and spatial reasoning. Classical location/POI recommendation ranks venues by check-in or rating signals~\citep{ye2011exploiting,zheng2017joint,menk2019tourism}; aspect-aware tourism adds review-extracted attributes~\citep{bauman2017aspect,liu2022tourism,li2024sabtr,baral2017pers}. Single-turn tourism QA over reviews~\citep{contractor2021poi} and travel-planning resources~\citep{li2025retail,fang2022dtcrskg,tang2024itinera,dietz2022tradeoff,dietz2020cityrec} cover one-shot QA or itinerary construction. RAG-augmented tourism~\citep{karlovic2025tourism,wei2025tourllm,park2025travel,vicente2026travel,cruz2023smarttur}, walkability/spatial-RAG~\citep{amendola2025walkrag,yu2025spatialrag,ni2025tprag}, and concurrent travel-CRS workshop work~\citep{fu2022trace,banerjee2026trace} move toward this space; \benchmark{} fills the remaining intersection in one benchmark. Table~\ref{tab:benchmark_comparison} positions it against the closest CRS resources.

\section{Conclusion}
\label{sec:conclusion}

\benchmark{} introduces a benchmark unit for accountable tourism CRS: multi-turn recommendation with review-span evidence and recovery after rejection. Across \numbaseline{} baselines, it exposes a Three-Competency Gap: LLM Zero-Shot leads R@1/recovery (0.533/0.988) but trails Dense on CGS (0.636 vs.\ 0.864), while retrievers and synthesis systems fail elsewhere. Accuracy-only leaderboards miss the core risk: a system can be fluent and often correct while weakly grounded, or well-cited while poor at selecting the right POI. Future CRS work should optimize the joint target: right recommendation, verifiable evidence, adaptive repair.

\textbf{Future directions.} \emph{(i) Joint-target inference}: staged retrieve-justify-repair architectures and joint-objective RL on the three-competency reward. \emph{(ii) Semantic-first grounding}: closing the fuzzy-vs.-entailment gap via NLI-aware decoding or post-generation citation editing without collapsing citation density. \emph{(iii) Spatial-reasoning architectures}: map-conditioned retrieval or geographically-pretrained LMs, since a proximity graph alone does not lift accuracy. \emph{(iv) Scope extensions}: controlled multi-LLM, multilingual extension, real-traveler dialogues, and a live KB with current pricing.

\providecommand{\nolinenumbers}{}\nolinenumbers
\clearpage
\bibliographystyle{plainnat}
\bibliography{references}

\clearpage

\appendix

\section{Metric and Baseline Catalogs}
\label{app:catalogs}

This appendix consolidates the two reference catalogs cited from the body: Table~\ref{tab:metrics} lists all \nummetric{} evaluation metrics across seven dimensions, and Table~\ref{tab:baselines} groups the \numbaseline{} baselines by training paradigm and retrieval signal.

\begin{table}[H]
  \caption{Complete evaluation metrics across seven dimensions (\nummetric{} metrics). Metrics marked with $\star$ are novel to this benchmark. Standard metrics (BLEU-4, ROUGE-L, Recall, MRR, TSR, T2R, Relevance, Informativeness, Natural Flow) follow conventional definitions; detailed formulas in Appendix~\ref{app:metrics_full}.}
  \label{tab:metrics}
  \centering
  \scriptsize
  \setlength{\tabcolsep}{3pt}
  \renewcommand{\arraystretch}{0.85}
  \begin{tabular}{llp{6cm}c}
    \toprule
    \textbf{Dimension} & \textbf{Metric} & \textbf{Definition} & \textbf{Novel?} \\
    \midrule
    \multirow{2}{*}{Text Quality} & BLEU-4 & Smoothed 4-gram precision (Chen \& Cherry) & \\
    & ROUGE-L & F1 of longest common subsequence & \\
    \midrule
    \multirow{3}{*}{Rec. Quality} & Recall@1 & $|\text{gold} \cap \text{top{-}1}| / |\text{gold}|$ & \\
    & Recall@3 & $|\text{gold} \cap \text{top{-}3}| / |\text{gold}|$ & \\
    & MRR & Mean reciprocal rank of gold POIs & \\
    \midrule
    \multirow{5}{*}{Grounding} & CGS (Composite) & $\text{GS} \times \min(1, \text{CD}/0.05) \times (0.5 + 0.5 \times \text{PC})$ & $\star$ \\
    & Grounding Score & Frac.\ of quotes with $\geq$80\% fuzzy match to source reviews & $\star$ \\
    & Entailment Gnd. & Frac.\ of claims semantically entailed by source reviews (NLI) & $\star$ \\
    & Citation Density & Verbatim citation tokens / total response tokens & $\star$ \\
    & Provenance Cov. & Frac.\ of aspect mentions backed by nearby citations & $\star$ \\
    \midrule
    \multirow{4}{*}{\makecell[l]{Practical\\Value}} & Spatial Coherence & Frac.\ of recommended POI pairs within 2\,km & $\star$ \\
    & Route Efficiency & Total nearest-neighbor path distance (km) & $\star$ \\
    & Price Alignment & Frac.\ of recommended POIs within user's stated budget & $\star$ \\
    & Itin.\ Diversity & Unique POI types / 3 (restaurant, hotel, attraction) & $\star$ \\
    \midrule
    \multirow{3}{*}{Dialogue} & TSR & Task success rate (any correct rec.\ in dialogue) & \\
    & T2R & Turns to first correct recommendation & \\
    & Rejection Recovery & Correct rec.\ rate after \texttt{reject\_and\_refine} & $\star$ \\
    \midrule
    \multirow{5}{*}{LLM Judge} & Relevance & Does response address user need? (1--5) & \\
    & Informativeness & Useful, specific, actionable? (1--5) & \\
    & Grounding Quality & Claims supported by evidence? (1--5) & $\star$ \\
    & Natural Flow & Conversational naturalness? (1--5) & \\
    & Justification & Well-reasoned recommendation explanations? (1--5) & $\star$ \\
    \midrule
    \multirow{3}{*}{\makecell[l]{Human\\Eval}} & Citation Precision & Frac.\ of quotes rated Supported (ALCE-style, App.\ \ref{app:human_eval}) & $\star$ \\
    & Informativeness (H) & Human Likert (1--5), useful and specific & \\
    & Naturalness (H) & Human Likert (1--5), conversational quality & \\
    \bottomrule
  \end{tabular}
\end{table}

\begin{table}[H]
  \caption{Baseline catalog, grouped by training paradigm. \textbf{Group}: \textsc{Non-LLM} = retrieval and template-based generation; \textsc{LLM} = \texttt{gpt-5.4-mini} with various augmentations. \textbf{Retrieval} type: S = sparse, D = dense, A = aspect, G = geographic, KB = structured attributes, P = persona; \textbf{State} = explicit dialogue state tracking. $^\dagger$Memory-Aug.\ is \emph{not} session-isolated; it reads persona-keyed preferences from prior dialogues sharing the same persona and is therefore reported separately (Appendix~\ref{app:limitations}); all competency tables (Tables~\ref{tab:main_results}--\ref{tab:stress}) use only the 13 session-isolated baselines.}
  \label{tab:baselines}
  \centering
  \scriptsize
  \setlength{\tabcolsep}{3pt}
  \renewcommand{\arraystretch}{0.85}
  \begin{tabular}{cllccc}
    \toprule
    \textbf{\#} & \textbf{Baseline} & \textbf{Key Capability} & \textbf{LLM} & \textbf{Retrieval} & \textbf{State} \\
    \midrule
    \multicolumn{6}{l}{\emph{Non-LLM group}} \\
    1 & Popularity & Heuristic ranking (no dialogue context) & & & \\
    2 & TF-IDF & \textbf{Term matching} & & S & \\
    3 & Aspect & \textbf{Multi-aspect decomposition} & & A & \\
    4 & Dense & \textbf{Semantic similarity (embedding)} & & D & \\
    5 & Spatial & \textbf{Location-aware ranking} & & S+G & \\
    6 & Hybrid-RRF & \textbf{Multi-signal fusion} & & S+A+D & \\
    7 & Itinerary & \textbf{Proximity graph + greedy TSP} & & S+G & \\
    8 & Knowledge-Enh. & \textbf{Attribute-constraint matching} & & S+KB & \\
    9 & Persona-Grounded & \textbf{Persona aspect weighting} & & A+P & \\
    \midrule
    \multicolumn{6}{l}{\emph{LLM group (\texttt{gpt-5.4-mini})}} \\
    10 & LLM Zero-Shot & Unaugmented LLM (reference) & \checkmark & & \\
    11 & DST & \textbf{Preference state tracking} & \checkmark & S & \checkmark \\
    12 & RAG-Citation & \textbf{Citation constraint} & \checkmark & S & \\
    13 & Multi-Review Synth. & \textbf{Aspect-consensus synthesis} & \checkmark & A & \\
    14 & Memory-Aug.$^\dagger$ & \textbf{Cross-session preference memory} & \checkmark & D & \checkmark \\
    \bottomrule
  \end{tabular}
\end{table}

\clearpage

\section{Limitations}
\label{app:limitations}

\benchmark{} is scoped to English Yelp data over 8 U.S.\ cities. The dialogue generator is \texttt{gpt-5.4-mini}; cross-family LLM Zero-Shot runs with Anthropic Haiku~4.5 and Qwen3.5-Flash (Table~\ref{tab:main_results}) confirm the LLM-family pattern is not a single-model artifact, and a fully controlled multi-LLM leaderboard is left for a camera-ready release. The 90-item ALCE-style human evaluation calibrates the automated grounding metrics rather than ranking systems; agreement statistics and an annotator-heterogeneity diagnostic are in Appendix~\ref{app:human_eval}. Memory-Aug.\ uses cross-session persona memory and is reported separately from the 13 session-isolated baselines.

\section{Dialogue Generation Pipeline: Full Diagram}
\label{app:pipeline_full}

\begin{figure}[h]
  \centering
  \includegraphics[width=\linewidth]{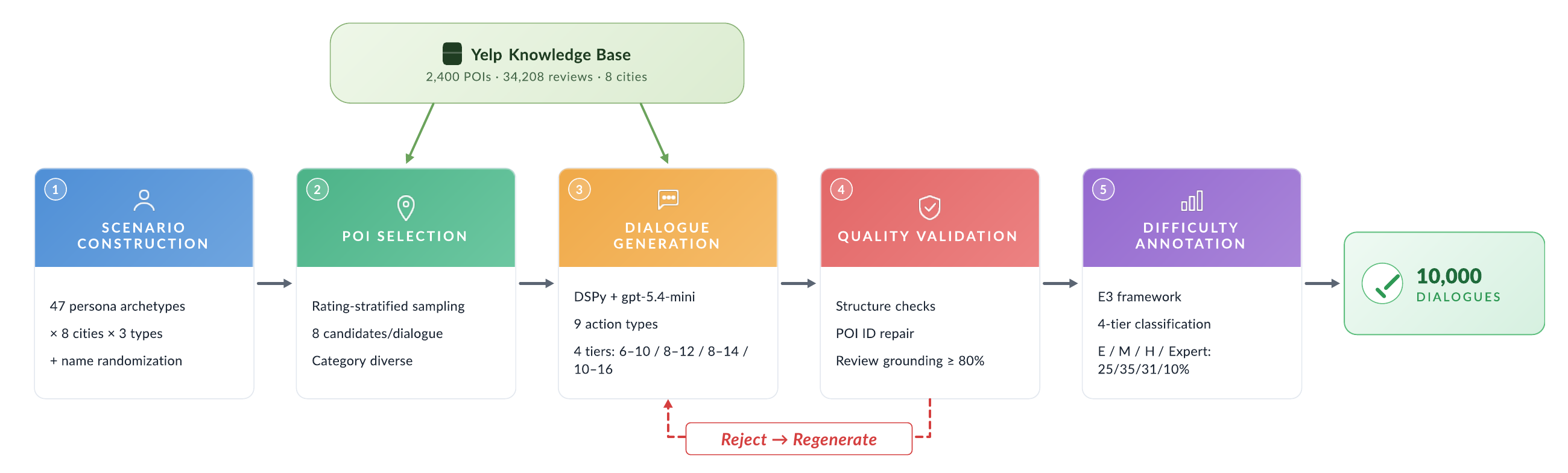}
  \caption{Five-stage dialogue generation pipeline. Stages feed sequentially, with a quality-validation gate that rejects structurally or factually flawed dialogues for regeneration.}
  \label{fig:pipeline}
\end{figure}

\section{Datasheet for \benchmark{}}
\label{app:datasheet}

Following the framework of \citet{gebru2021datasheets}, we provide a datasheet for \benchmark{}.

\paragraph{Motivation.}
\textit{For what purpose was the dataset created?} To enable research on evidence-grounded conversational recommender systems in the tourism domain.
\textit{Who created the dataset?} The authors of this paper.
\textit{Who funded the creation?} [Anonymized for review].

\paragraph{Composition.}
\textit{What do the instances represent?} Multi-turn dialogues between a simulated tourist (user) and a recommender system (system), grounded in real Yelp POI data and reviews.
\textit{How many instances are there?} \numdiag{} dialogues, \numpoi{} POIs, \numreview{} reviews.
\textit{Is there a label or target?} Each system turn is annotated with action labels, referenced POI IDs, and review citation IDs.

\paragraph{Collection Process.}
\textit{How was the data acquired?} Dialogues were synthetically generated via DSPy with \texttt{gpt-5.4-mini}, grounded in the Yelp Academic Dataset. POI data is from the Yelp Open Dataset under its terms of use.
\textit{Was any preprocessing applied?} Reviews were truncated to 500 characters. POIs were filtered to require a minimum of 3 reviews.

\paragraph{Uses.}
\textit{Has the dataset been used already?} Only for the experiments in this paper.
\textit{What tasks could it be used for?} Conversational recommendation, review-grounded generation, dialogue state tracking, evidence attribution evaluation.

\paragraph{Distribution.}
\textit{How is the dataset distributed?} The benchmark, all baseline implementations, and the evaluation toolkit are released at \url{https://anonymous.4open.science/r/TRACE-benchmark}. Mirroring the main Release paragraph (\S\ref{sec:construction_detail}): the synthetic content we authored, namely (a) the \numdiag{} dialogues we generated (dialogue structure, non-quote turn text, action labels, grounding annotations, citation-pointer metadata as \texttt{(business\_id, R\#)} tuples) and (b) the \numbaseline{} baseline implementations and evaluation toolkit, is released under \textbf{CC BY 4.0}. Verbatim Yelp review spans are \emph{not} redistributed: in released dialogue text they appear as \texttt{[Q:N]} placeholders backed by positional metadata (review ID plus character offsets), and the full text is hydrated locally from a user-obtained Yelp Open Dataset bundle under Yelp's academic-use terms (non-commercial, non-sublicensable, revocable; \S3, \S4.A, \S4.H, \S10 of the Feb 2021 agreement). Yelp-derived POI metadata (name, address, lat/lon, categories, stars, price) is included under the \S4.E academic-publication carve-out; the \texttt{business\_id} and \texttt{review\_id} values appear as bare factual identifiers. The full \numreview{} review corpus is not redistributed. A blanket MIT or Apache-2.0 release is \emph{not} legally available because Yelp's User Agreement is non-sublicensable and non-commercial, which conflicts with permissive open-source license grants.

\paragraph{Maintenance.}
\textit{Who will maintain the dataset?} The authors, with plans for community-driven extensions.

\section{Practical Value Metrics: Full Table}
\label{app:practical_value}

Table~\ref{tab:spatial} reports the practical-value diagnostics (spatial coherence, price alignment, itinerary diversity, raw Grounding Score) for all 13 session-isolated baselines on the closed-set protocol. Itinerary planning is the only baseline that meaningfully improves spatial coherence; price alignment is high across all baselines because most user-stated budgets are loose; itinerary diversity is at the floor for all but Itinerary, which spreads recommendations across POI types by construction.

\begin{table}[H]
  \caption{Practical value metrics (closed-set, all 13 session-isolated baselines). \textbf{Spatial Coherence}: fraction of POI pairs within 2\,km on recommend/compare turns. \textbf{Price Alignment}: budget compliance rate when the user states a price ceiling. \textbf{Itin.\ Div.}: unique POI types / 3. GS = raw Grounding Score (see Table~\ref{tab:main_results} for CGS). Best per column in \textbf{bold}, second best \underline{underlined}.}
  \label{tab:spatial}
  \centering
  \small
  \begin{tabular}{lcccc}
    \toprule
    \textbf{Baseline} & \textbf{Spatial Coh.}$\uparrow$ & \textbf{Price Align.}$\uparrow$ & \textbf{Itin. Div.}$\uparrow$ & \textbf{GS} \\
    \midrule
    Popularity         & 0.174 & \underline{0.974} & 0.333 & 0.998 \\
    TF-IDF             & 0.177 & 0.957 & 0.333 & 0.998 \\
    Aspect             & 0.175 & 0.953 & 0.333 & 0.998 \\
    Dense              & 0.186 & 0.966 & 0.333 & 0.997 \\
    Spatial            & \underline{0.195} & 0.958 & 0.333 & 0.997 \\
    Hybrid-RRF         & 0.192 & 0.956 & 0.333 & 0.997 \\
    Itinerary          & \textbf{0.640} & 0.953 & \textbf{0.525} & 0.997 \\
    Knowledge-Enh.     & 0.165 & 0.968 & 0.333 & \textbf{0.999} \\
    Persona-Ground.    & 0.173 & 0.944 & 0.333 & \textbf{0.999} \\
    \midrule
    LLM Zero-Shot      & \underline{0.195} & 0.955 & 0.333 & 0.981 \\
    DST                & 0.189 & \textbf{0.976} & 0.333 & 0.982 \\
    RAG-Citation       & 0.193 & 0.954 & 0.333 & 0.994 \\
    Multi-Rev.\ Synth. & 0.186 & 0.960 & 0.333 & 0.998 \\
    \bottomrule
  \end{tabular}
\end{table}

\section{Baseline Implementation Details}
\label{app:baselines}

\paragraph{PopularityBaseline.}
Ranks POIs by $(\text{stars}, \text{review\_count})$ in descending order, ignoring all dialogue context. Uses fixed template responses per action type. Serves as a lower bound and fallback for other baselines when they fail.

\paragraph{TFIDFBaseline.}
Builds a per-POI document by concatenating the POI name, categories, and all review texts. Computes TF-IDF with logarithmic term frequency: $\text{tf}(t,d) = 1 + \log(\text{count}(t,d))$ and smoothed inverse document frequency: $\text{idf}(t) = \log\frac{n+1}{\text{df}(t)+1} + 1$. Removes 80 stop words. The query is the concatenation of all user turns. Caches the TF-IDF index by (city, POI type) for efficiency.

\paragraph{LLMZeroShotBaseline.}
Provides dialogue history and formatted POI cards to \texttt{gpt-5.4-mini} with a one-sentence action instruction. Does not use dialogue guidelines or structured constraints. Falls back to PopularityBaseline on LLM failure.

\paragraph{DSTBaseline.}
Two-stage per-turn pipeline: (1)~DSPy extracts a \texttt{TourismDialogueState} delta (price range, cuisine preferences/avoidances, ambiance, noise level, required features, location constraint, rejected/accepted POIs) from the latest user turn; (2)~merges into a cumulative state. The state filters candidate POIs (price $\leq$ stated range, cuisine positive/negative filters, noise level compatibility, feature requirements) before TF-IDF ranking. An LLM generates the final response conditioned on the structured state.

\paragraph{RAGCitationBaseline.}
Retrieves top-10 review sentences by Jaccard similarity to the user query. Formats evidence as numbered citations with POI names and star ratings. The LLM must include at least one verbatim quote ($\geq$10 tokens); up to 2 retries if citation check fails.

\paragraph{AspectRetrievalBaseline.}
Five aspect categories: food, service, ambiance, value, location. Per-review aspect sentiment via keyword-based detection. POI scoring: $\text{score} = \sum_a w_a \cdot s_a + 0.1 \cdot \text{stars}$, where $w_a = \min(\text{hits}_a / 3, 1.0)$ is the query aspect weight and $s_a$ is the mean aspect sentiment.

\paragraph{DenseRetrievalBaseline.}
Uses \texttt{all-MiniLM-L6-v2} sentence embeddings. POI documents: name + categories + top-3 reviews (200 chars each). Negative preference: subtracts rejected POI embeddings with weight 0.5; hard penalty of $-1.0$ for explicitly rejected POIs.

\paragraph{SpatialBaseline.}
Blended ranking: $\text{score} = 0.7 \cdot \text{tf\_score} + 0.3 \cdot \text{spatial\_score}$, where spatial score $= 1 - \text{dist}/\text{max\_dist}$ using haversine distance. Anchor detection: keyword matching for location references in dialogue, fallback to city centroid.

\paragraph{HybridRRFBaseline.}
Reciprocal Rank Fusion over four signals: $\text{score}(d) = \sum_{r \in \text{rankers}} \frac{1}{k + \text{rank}_r(d)}$, with $k=60$. Rankers: TF-IDF, Aspect, Dense, and structured attributes (stars + price). Falls back to 3-way fusion if dense retrieval is unavailable.

\paragraph{MultiReviewSynthesisBaseline.}
Clusters review sentences by aspect. Computes per-aspect consensus: $c_a = (n_{\text{pos}} - n_{\text{neg}}) / n_{\text{total}} \in [-1, 1]$. Generates aspect-focused responses highlighting consensus aspects.

\paragraph{MemoryAugmentedBaseline.}
Maintains a persistent \texttt{PreferenceMemory} across dialogue sessions. Preference retrieval uses cosine similarity with exponential time decay: $\text{decay} = \exp(-\ln 2 \cdot \text{hours\_ago} / 168)$, giving a half-life of one week. Memory persists across session resets.

\paragraph{ItineraryBaseline.}
Constructs a proximity graph: POIs within 500\,m are connected, with same-type edges excluded. Visit ordering via greedy nearest-neighbor TSP from the first coordinate-valid POI. Responses include distance annotations (meters if $<$1\,km, otherwise km).

\section{Prompt Templates}
\label{app:prompts}

The dialogue generation prompt (V2) includes the following key instructions:

\begin{itemize}[nosep,leftmargin=*]
  \item Generate 6--12 turns with strict user/system alternation
  \item User initiates (\texttt{greet\_and\_seek}), system concludes (\texttt{farewell})
  \item At least 1 \texttt{ask\_preference} action and 1 \texttt{reject\_and\_refine}
  \item Recommend $\geq$3 distinct POIs with $\geq$1 comparison
  \item System responses MUST include EXACT verbatim review quotes (substrings allowed, NO paraphrasing) with \texttt{[Rn]} citation labels
  \item JSON output format with \texttt{turn\_id}, \texttt{role}, \texttt{text}, \texttt{action}, \texttt{referenced\_poi\_ids}, \texttt{referenced\_review\_ids}
\end{itemize}

\section{Dialogue Examples}
\label{app:examples}

The full dataset (\texttt{crs\_dataset.json}, \numdiag{} dialogues) is released in the supplementary material; each dialogue is stored as a JSON object with fields \texttt{dialogue\_id}, \texttt{scenario} (persona/city/POI type), \texttt{candidate\_poi\_ids}, \texttt{turns} (each with \texttt{role}, \texttt{text}, \texttt{action}, \texttt{referenced\_poi\_ids}, \texttt{referenced\_review\_ids}), and \texttt{metadata} (difficulty tier, generation time). Representative dialogues span all 8 cities (Indianapolis, Nashville, New Orleans, Philadelphia, Reno, Saint Louis, Tampa, Tucson) $\times$ 3 POI types $\times$ 47 personas; recommended sample IDs: \texttt{crs\_yelp\_08506} (Tampa restaurant, date\_night, Hard), \texttt{crs\_yelp\_02669} (Nashville hotel, layover\_traveler, Medium), \texttt{crs\_yelp\_04339} (New Orleans attraction, family, Easy). \emph{Per-POI cards} (the released file \texttt{poi\_knowledge\_base.json}) contain only the citation pointers \texttt{(business\_id, R\#)} and the short verbatim spans cited in the released dialogues; the full review text and structured POI attributes must be re-derived locally from a user-obtained Yelp Open Dataset bundle (Datasheet, Appendix~\ref{app:datasheet}).

\section{Persona Templates}
\label{app:personas}

The full set of 47 persona archetypes (verified from \texttt{crs\_dataset.json} \texttt{scenario.persona\_type}, sorted alphabetically):

\noindent\texttt{accessibility\_needs, architecture\_buff, brunch\_enthusiast, budget\_traveler, business, celebration, conference\_attendee, couple, date\_night, dietary\_restricted, digital\_nomad, eco\_tourist, eco\_traveler, event\_group, event\_planner, extended\_stay, family, fitness\_traveler, food\_explorer, foodie\_explorer, friends\_group, group\_event, health\_conscious, history\_buff, international\_tourist, late\_night, layover\_traveler, local\_experience, luxury\_couple, luxury\_traveler, medical\_traveler, music\_lover, nightlife\_seeker, pet\_owner, relocation, remote\_worker, road\_tripper, senior, solo\_traveler, sports\_fan, street\_art\_seeker, thrill\_seeker, traveling\_nurse, urban\_explorer, volunteer\_tourist, wellness\_seeker, wine\_enthusiast}.

Each persona archetype is paired with travel contexts (\eg, first visit, anniversary weekend, business trip, layover, relocation scouting). Dialogues cycle through all 47 archetypes across 8 cities and 3 POI types; POI-type-specific persona pools (38 per type) ensure contextually appropriate scenarios (\eg, hotel personas include conference attendees and layover travelers). A randomization layer further diversifies persona descriptions by varying names, ages, and appending detail modifiers (\eg, ``who is vegetarian,'' ``who uses a wheelchair'') with 30\% probability.

\section{Metrics: Detailed Definitions}
\label{app:metrics_full}

This appendix expands the standard metrics referenced in Section~\ref{sec:evaluation}. The novel grounding, practical-value, and human-evaluation metrics are described in the main text.

\paragraph{Text Quality.} \textbf{BLEU-4} is computed with the Chen \& Cherry smoothing on tokenized response/reference pairs at the turn level. \textbf{ROUGE-L} reports the F1 of the longest common subsequence between the system response and the reference dialogue's response for the same turn. Both are widely used surface-similarity metrics; we report them for comparability with prior CRS work even though neither captures grounding fidelity or recommendation correctness.

\paragraph{Recommendation Quality.} \textbf{Recall@1} and \textbf{Recall@3} compute the fraction of gold POIs (the set of POIs recommended in the reference dialogue) that appear in the system's top-1 / top-3 prediction list for the current turn. \textbf{MRR} (Mean Reciprocal Rank) reports the mean of $1/\text{rank}$ across recommend/compare turns, where rank is the position of the first gold POI in the system's ranked list (1-indexed; defaults to 0 if no gold POI is in the list).

\paragraph{Dialogue.} \textbf{TSR} (Task Success Rate) is binary at the dialogue level: 1.0 if any system turn recommends a gold POI by exact business-ID match, 0.0 otherwise. \textbf{T2R} (Turns to first correct Recommendation) counts system turns with \texttt{recommend} or \texttt{compare} actions until the first gold-POI hit; lower is better. \textbf{Rejection Recovery Rate} (novel) is the fraction of \texttt{reject\_and\_refine} episodes whose immediately following \texttt{recommend}/\texttt{compare} turn hits a gold POI.

\paragraph{LLM-as-Judge.} An LLM judge rates each response on five 1--5 Likert scales: \textbf{Relevance} (does the response address the user's stated need?), \textbf{Informativeness} (is it specific and actionable?), \textbf{Grounding Quality} (are claims supported by evidence?, novel), \textbf{Natural Flow} (conversational naturalness), and \textbf{Justification} (well-reasoned recommendation explanations?, novel). The headline LLM-as-Judge ratings (Appendix~\ref{app:llm_judge_full}) use Gemini~2.5~Flash. We additionally report a narrow cross-judge agreement check (GPT-4o vs.\ Claude Sonnet 4) on a 200-dialogue subset restricted to the 6 non-LLM baselines in Appendix~\ref{app:cross_judge}; that two-judge agreement is therefore complementary to the Gemini ratings, not a re-scoring of them, and does not cover LLM baselines.

\paragraph{Provenance Coverage (PC) algorithm.}
PC quantifies whether a response's \emph{aspect mentions} (claims about specific POI attributes) are accompanied by nearby citations. Algorithm~\ref{alg:pc} gives the exact procedure used in the released code. Inputs: a system response $r$ (text plus parsed citation labels with positions), a referenced-POI catalog $K$ (each POI carries a fixed set of categorical and amenity attributes). The aspect inventory $\mathcal{A}$ is a hand-curated 41-term seed list spanning four families: \emph{food/menu items} (e.g., \texttt{seafood}, \texttt{vegan}, \texttt{breakfast}), \emph{ambience/style} (e.g., \texttt{quiet}, \texttt{cozy}, \texttt{romantic}), \emph{logistics} (e.g., \texttt{parking}, \texttt{wifi}, \texttt{outdoor seating}), and \emph{value/price} (e.g., \texttt{cheap}, \texttt{upscale}, \texttt{family-friendly}); the full list is in the released \texttt{src/tqa/crs/metrics.py}. We use case-insensitive whole-word match (\texttt{re.findall} with word boundaries), no morphological expansion or synonym resolution. The \emph{citation-neighborhood window} $w$ is 80 characters by default (tuneable; sensitivity reported in Appendix~\ref{app:sensitivity}). A mention at character offset $p$ is counted as \emph{covered} if some citation label \texttt{[Rn]} appears within $[p-w, p+w]$. Tie handling: if multiple aspect terms overlap (e.g., ``family-friendly cafe'' matches both \texttt{family-friendly} and \texttt{cafe}), each is counted once; duplicate mentions of the same term in the same response are merged. PC is the per-response fraction of covered mentions, averaged over the dialogue's recommend/compare turns; turns with zero aspect mentions contribute PC = 1.0 (vacuously true) and are excluded from the per-turn denominator only when they contain no recommendation either.

\begin{algorithm}[h]
\caption{Provenance Coverage (PC)}
\label{alg:pc}
\begin{algorithmic}[1]
\Require Response text $r$, citation positions $C = \{(\ell_i, p_i)\}$, aspect inventory $\mathcal{A}$, window $w$
\State $M \gets \emptyset$ \Comment{set of (aspect, position) mentions}
\For{aspect $a \in \mathcal{A}$}
    \For{$p \in \mathrm{find\_word\_boundaries}(r, a, \mathrm{case\_insensitive})$}
        \State $M \gets M \cup \{(a, p)\}$
    \EndFor
\EndFor
\If{$|M| = 0$} \Return $1.0$ \Comment{vacuous truth: no claims to ground} \EndIf
\State $\mathit{covered} \gets 0$
\For{$(a, p) \in M$}
    \If{$\exists\,(\ell_i, p_i) \in C$ with $|p_i - p| \leq w$}
        \State $\mathit{covered} \gets \mathit{covered} + 1$
    \EndIf
\EndFor
\State \Return $\mathit{covered} \,/\, |M|$
\end{algorithmic}
\end{algorithm}

\section{Human Evaluation: Details}
\label{app:human_eval}

This appendix supplements Section~\ref{sec:human_eval_main}. It documents (a) the verbatim instructions and screenshots given to participants, (b) per-baseline counts and the error-category breakdown for Task 1, (c) an annotator-heterogeneity diagnostic for IAA, and (d) an earlier pilot study.

\subsection{Verbatim Instructions Given to Participants}
\label{app:human_eval_instructions}

\begin{figure}[t]
  \centering
  \includegraphics[width=0.78\linewidth]{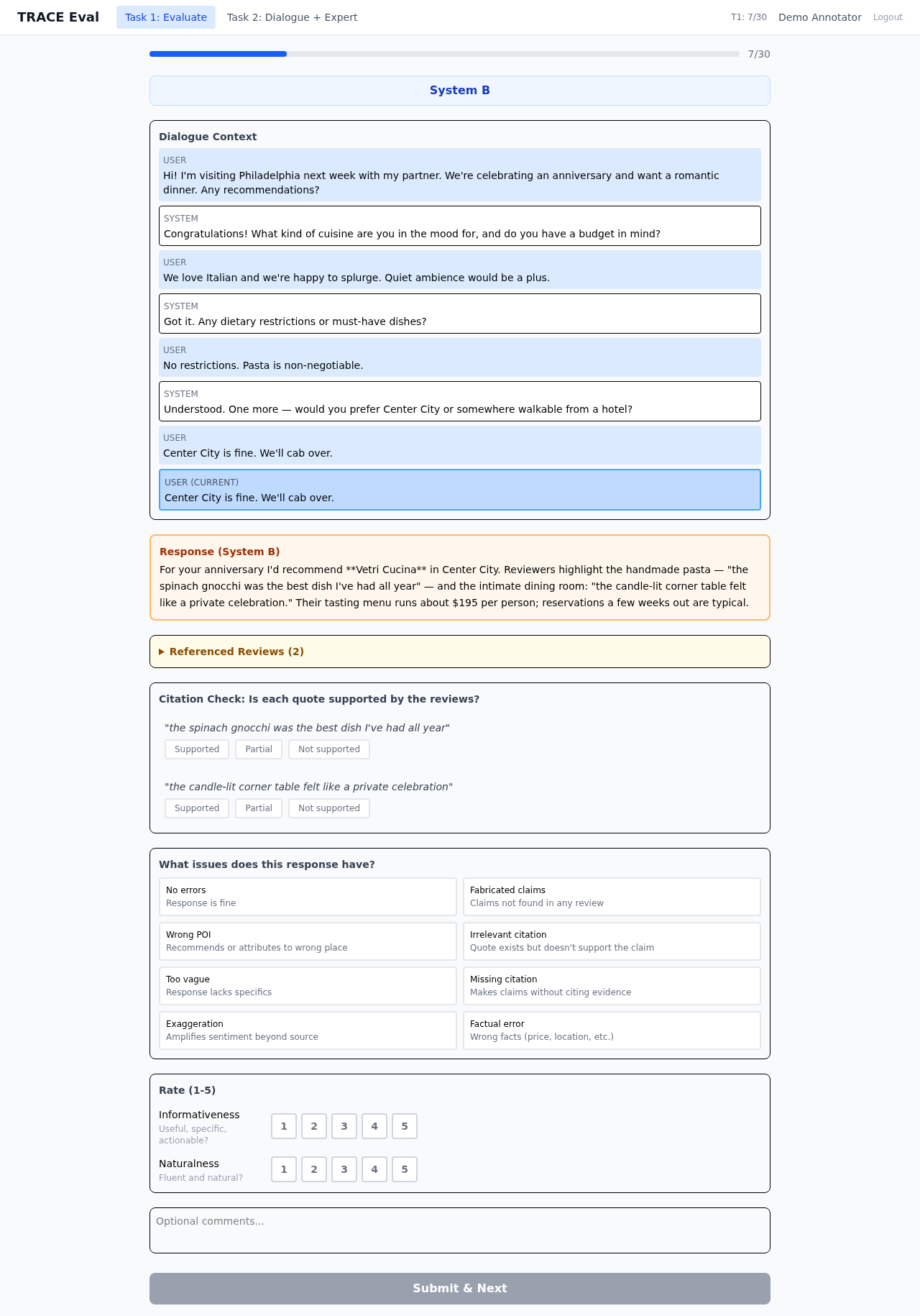}
  \caption{\textbf{Task 1: baseline response rating page.} The annotator sees the full dialogue context (top), the response to evaluate (orange box), the cited reviews (collapsible), the per-quote Citation Check (Supported / Partial / Not supported), the 8-way error checklist, and two 5-point Likert scales (Informativeness, Naturalness). The system label (System~A/B/C) is randomized per item; annotators do not know which baseline produced the response. The example shown is illustrative; production tasks were drawn from the 90-item evaluation set.}
  \label{fig:annot_task1}
\end{figure}

\begin{figure}[t]
  \centering
  \includegraphics[width=0.78\linewidth]{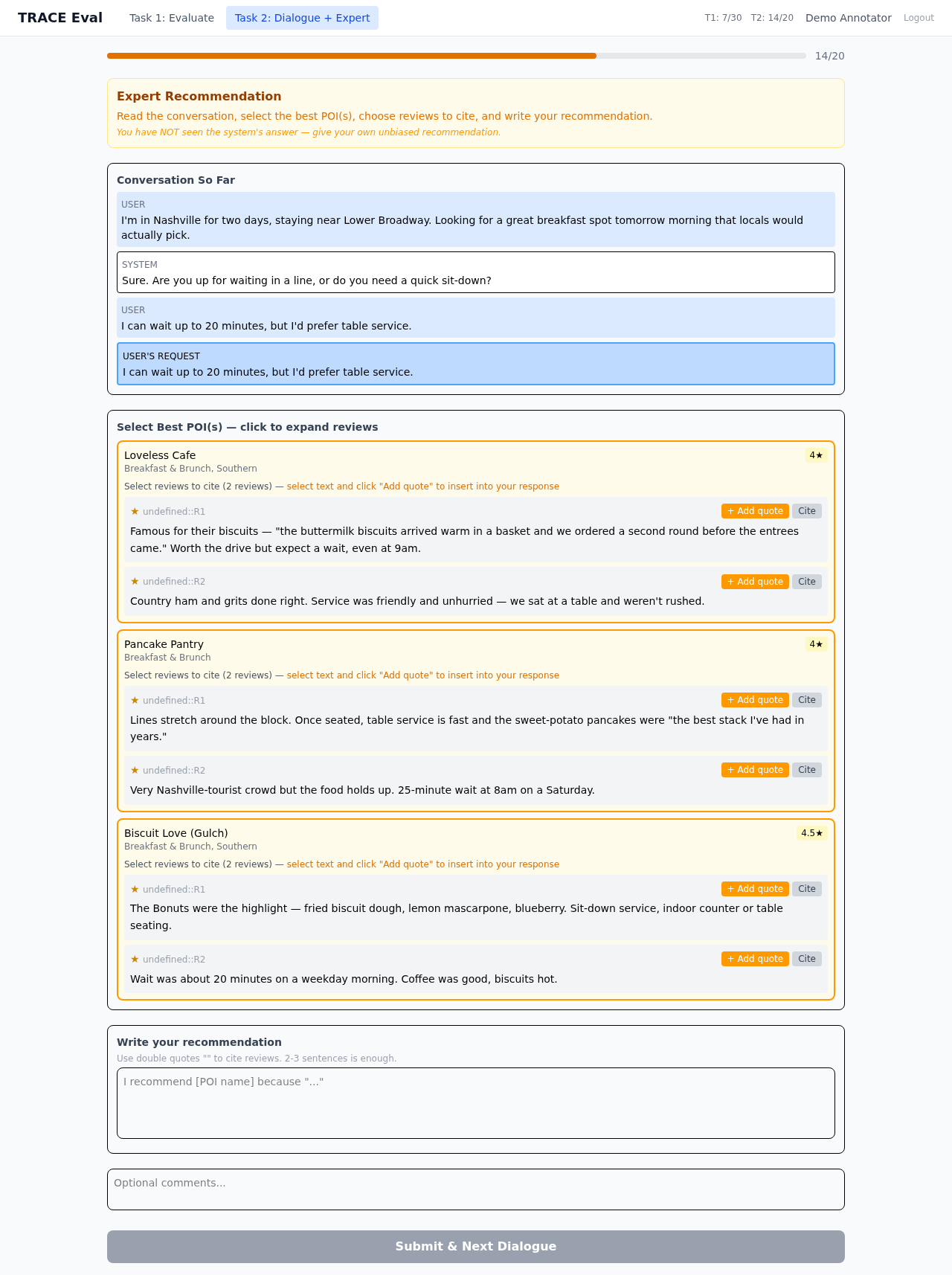}
  \caption{\textbf{Task 2: expert recommendation page} with all candidate POIs expanded to show their reviews. The annotator sees the conversation so far (with the user's current request highlighted in blue), 8 candidate POIs (3 shown here for the illustrative example) each with reviews, and a free-text response box. Annotators select review spans and click the orange `+~Add quote' button, which inserts a verbatim citation into the response and marks the POI as selected. Annotators do NOT see the system's actual answer for that dialogue, eliminating anchoring bias.}
  \label{fig:annot_task2_expanded}
\end{figure}

Annotators received the briefing reproduced verbatim below before starting the task. Each annotator also went through a 15-minute walk-through of the web tool with one of the authors, in which the screens shown in Figures~\ref{fig:annot_task1}--\ref{fig:annot_task2_expanded} were demonstrated on three calibration items. The full annotation tool is open-sourced together with the benchmark.

\paragraph{Overview (shown on the first page).}
``You will evaluate responses from AI tourism recommendation systems. The systems recommend restaurants, hotels, and attractions to travelers based on their preferences, citing real user reviews as evidence. There are 2 tasks. Each annotator completes both tasks: Task~1 (Rate AI system responses, 90 items, $\sim$1~hour), Task~2 (Write your own recommendation, 20 items, $\sim$30~minutes). Total $\sim$1.5 hours.''

\paragraph{Task 1 instructions (verbatim).}
``\emph{What you see.} For each item, you will see (1) a dialogue context between a tourist (User) and a recommendation system (System); read the full conversation to understand what the user wants. (2) A system label `System A', `System B', or `System C'. You do NOT know which AI model produced it. (3) The response to evaluate, highlighted in orange. (4) Referenced reviews --- the actual user reviews that the system cited; click to expand.

\emph{Part A. Citation Check (per quote).} Each quoted phrase (in double quotes) from the response is extracted automatically. For each quote, judge: \textbf{Supported} --- the quote accurately appears in the referenced reviews and supports the claim being made (the exact text or a very close paraphrase exists, AND it is used correctly). \textbf{Partial} --- the quote exists in reviews but is taken out of context, or only partially supports the claim (the text is real but misrepresents the reviewer's intent, or the claim goes beyond what the quote says). \textbf{Not supported} --- the quote cannot be found in any referenced review, or is fabricated. \emph{How to check}: expand `Referenced Reviews' and search for the quoted text. If you find the exact text or a very close match, it is Supported. If you find similar but not matching text, it is Partial. If you cannot find it at all, it is Not supported.

\emph{Part B. Error Checklist.} Select ALL errors you find in the response. If the response has no problems, select `No errors'. Categories: \textbf{No errors} (response is fine; select ONLY if no other error applies); \textbf{Fabricated claims} (claims not found in any review, e.g.\ `This restaurant has a Michelin star' when no review mentions this); \textbf{Wrong POI} (recommends or attributes info to wrong place, e.g.\ talks about Restaurant~A but quotes reviews from Restaurant~B); \textbf{Irrelevant citation} (quote exists but does not support the claim, e.g.\ quoting a review about desserts when discussing parking); \textbf{Too vague} (response lacks specifics, e.g.\ `This is a great restaurant' with no details); \textbf{Missing citation} (makes specific claims without citing evidence, e.g.\ `The pasta is excellent' with no review quoted); \textbf{Exaggeration} (amplifies sentiment beyond source review, e.g.\ review says `decent food', system says `extraordinary cuisine'); \textbf{Factual error} (wrong facts: price, location, etc.).

\emph{Part C. Likert Ratings (1--5).} \textbf{Informativeness}: 1 = generic, no useful details; 3 = some useful info but incomplete; 5 = specific, actionable, helps user decide. \textbf{Naturalness}: 1 = robotic, awkward, or template-like; 3 = readable but clearly AI-generated; 5 = fluent, sounds like a knowledgeable travel agent.

\emph{Tips.} Read the FULL dialogue context before evaluating. A response can have high naturalness but fabricated citations (or vice versa); rate each dimension independently. `No errors' and error categories are mutually exclusive --- if you select `No errors', do not select anything else. You do not need to verify every factual claim --- focus on the quoted citations. If unsure whether a quote is supported, default to Partial.''

\paragraph{Task 2 instructions (verbatim).}
``For each dialogue you write your own recommendation. You see (1) the conversation up to the point where the system would normally recommend; (2) the user's current request, highlighted in blue; (3) 8 candidate POIs, each with name, rating, categories, price range, and reviews; click a POI to expand its reviews. \textbf{You do NOT see the system's actual answer} --- this is intentional; we want your unbiased recommendation.

\emph{Your task.} (1) Select the best POI(s) --- click on the POI name to expand its reviews; you can select 1--3 POIs. (2) Add quotes from reviews --- in the expanded reviews, use your mouse to select (highlight) a phrase, then click the `+ Add quote' button. The system will automatically generate a recommendation sentence (\texttt{I recommend [POI name] because ``[your selected text]''}), mark the POI as selected, and mark the review as cited; additional quotes are appended automatically. (3) Edit the recommendation --- review the auto-generated text in the text box; add context, adjust wording, or add more quotes; 2--3 sentences is enough.

\emph{Guidelines for writing.} Recommend based on what the USER asked for, not your personal preferences. Quotes should be verbatim from reviews (the `Add quote' button handles this for you). Be specific: mention what makes this POI a good fit for THIS user. 2--3 sentences is enough.''

\paragraph{General rules (verbatim).}
``(1) Be consistent. Apply the same standards across all items. (2) Use the full scale. Do not bunch everything in 3--4. Use 1 and 5 when they are genuinely warranted. (3) Take breaks. This is $\sim$1.5 hours of focused work. (4) Do not look up POIs. Judge based only on the information shown to you. Do not Google the restaurants. (5) Optional comments. Use the comment box if something is unusual or you want to explain a rating.''

\subsection{Per-Baseline Counts and Error Breakdown}
\label{app:human_eval_counts}

Annotators evaluate 90 system responses (30 dialogues $\times$ 3 baselines: TF-IDF, RAG-Citation, LLM Zero-Shot). Each response is assessed on (a)~per-quote citation accuracy (Supported / Partial / Not Supported), (b)~an error checklist (8 categories: \emph{fabrication}, \emph{wrong\_poi}, \emph{irrelevant\_citation}, \emph{too\_vague}, \emph{missing\_citation}, \emph{exaggeration}, \emph{factual\_error}, \emph{no\_errors}), and (c)~two 5-point Likert scales (informativeness, naturalness). Five trained research assistants completed the task independently on the same 90 items for IAA. Headline numbers in Section~\ref{sec:human_eval_main}; full per-baseline counts in Table~\ref{tab:human_baseline_ratings}.

\begin{table}[h]
  \caption{Human evaluation of baseline responses (5 annotators, 30 dialogues $\times$ 3 baselines = 90 items). Likert columns are 5-annotator means; citation columns are majority-vote counts; Citation Precision = fraction of quotes with majority label \emph{Supported}.}
  \label{tab:human_baseline_ratings}
  \centering \small
  \begin{tabular}{lcccccc}
    \toprule
    & \multicolumn{2}{c}{\textbf{Likert (1--5, mean)}} & \multicolumn{3}{c}{\textbf{Citation Check (majority)}} & \\
    \textbf{Baseline} & \textbf{Info.} & \textbf{Nat.} & \textbf{Supp.} & \textbf{Part.} & \textbf{Not S.} & \textbf{Cite.Prec.} \\
    \midrule
    TF-IDF        & 1.55 & 2.33 & 40 &  4 &  0 & 0.909 \\
    RAG-Citation  & 2.91 & 3.85 & 63 &  0 &  2 & \textbf{0.969} \\
    LLM Zero-Shot & \textbf{3.55} & \textbf{3.97} & 151 & 12 & 32 & 0.774 \\
    \bottomrule
  \end{tabular}
\end{table}

\paragraph{Task 1: Error-category breakdown.}
The most common errors across all 90 items are: \emph{too vague} (66.7\%), \emph{irrelevant citation} (61.1\%), \emph{wrong POI} (32.2\%), and \emph{factual error} (23.3\%). Only 5.6\% of responses were judged error-free. TF-IDF triggers \emph{too vague} in 100\% of cases (its template responses lack specificity by construction), while LLM Zero-Shot's primary weakness is \emph{irrelevant citation} (67\%), where quoted text exists but does not support the specific user claim. Error-category $\alpha{=}0.31$ reflects the multi-label nature of the task (annotators often agree on presence of errors but differ on which specific categories to check).

\paragraph{Annotator Heterogeneity (Leave-One-Out).}
We report all headline agreement values using all five annotators. As a diagnostic, \texttt{scripts/iaa\_leave\_out\_analysis.py} recomputes Krippendorff's $\alpha$ after dropping annotators from the shared 90-item overlap and summarizes per-annotator rating behavior. Annotator IDs are anonymized as A1--A5.

\begin{table}[h]
  \caption{Per-annotator diagnostics on the 90-item Task 1 overlap. Means are 5-point Likert averages; deviations are mean absolute deviations from the item-level group median; flag rate is the fraction of items where the annotator marked any non-\emph{no\_errors} category; avg.~cats is the mean number of error categories selected per item. A5 is a systematic soft rater: highest Likert means, roughly twice the median deviation of the other annotators, lowest error-flagging rate, and lowest error-category density.}
  \label{tab:human_annotator_diagnostics}
  \centering \small
  \begin{tabular}{lcccccc}
    \toprule
    \textbf{Annotator} & \textbf{Info.\ mean} & \textbf{Nat.\ mean} & \textbf{Info.\ dev.} & \textbf{Nat.\ dev.} & \textbf{Flag rate} & \textbf{Avg.\ cats} \\
    \midrule
    A1 & 2.60 & 3.31 & 0.49 & 0.37 & 76.7\% & 1.59 \\
    A2 & 2.36 & 2.96 & 0.20 & 0.34 & 92.2\% & 2.03 \\
    A3 & 2.80 & 3.67 & 0.60 & 0.50 & 76.7\% & 1.72 \\
    A4 & 2.16 & 2.96 & 0.29 & 0.41 & 94.4\% & 2.16 \\
    A5 & \textbf{3.42} & \textbf{4.04} & \textbf{1.02} & \textbf{0.90} & \textbf{61.1\%} & \textbf{0.79} \\
    \bottomrule
  \end{tabular}
\end{table}

\begin{table}[h]
  \caption{Leave-one-out Krippendorff's $\alpha$ diagnostics. Baseline all-five agreement is $\alpha_{\text{info}}{=}0.578$, $\alpha_{\text{nat}}{=}0.423$, $\alpha_{\text{cite}}{=}0.319$, $\alpha_{\text{err}}{=}0.314$ (mean $=0.408$). Mean / $\Delta$ is the mean across the four dimensions and its change relative to all five.}
  \label{tab:human_leave_one_out}
  \centering \small
  \begin{tabular}{lccccc}
    \toprule
    \textbf{Dropped} & $\alpha_{\text{info}}$ & $\alpha_{\text{nat}}$ & $\alpha_{\text{cite}}$ & $\alpha_{\text{err}}$ & \textbf{Mean / $\Delta$} \\
    \midrule
    A1 & 0.578 & 0.395 & 0.253 & 0.322 & 0.387 / $-0.021$ \\
    A2 & 0.524 & 0.373 & 0.312 & 0.252 & 0.365 / $-0.043$ \\
    A3 & 0.605 & 0.467 & 0.357 & 0.279 & 0.427 / $+0.019$ \\
    A4 & 0.547 & 0.362 & 0.292 & 0.287 & 0.372 / $-0.036$ \\
    A5 & \textbf{0.625} & \textbf{0.505} & \textbf{0.366} & \textbf{0.418} & \textbf{0.479 / $+0.070$} \\
    \bottomrule
  \end{tabular}
\end{table}

\begin{table}[h]
  \caption{Leave-two-out diagnostics, top five pairs by mean $\alpha$. These are diagnostic only and are not used for headline reporting.}
  \label{tab:human_leave_two_out}
  \centering \small
  \begin{tabular}{lccccc}
    \toprule
    \textbf{Dropped pair} & $\alpha_{\text{info}}$ & $\alpha_{\text{nat}}$ & $\alpha_{\text{cite}}$ & $\alpha_{\text{err}}$ & \textbf{Mean / $\Delta$} \\
    \midrule
    A3 + A5 & \textbf{0.732} & \textbf{0.709} & \textbf{0.430} & 0.422 & \textbf{0.573 / $+0.165$} \\
    A1 + A5 & 0.645 & 0.480 & 0.289 & \textbf{0.475} & 0.472 / $+0.064$ \\
    A4 + A5 & 0.569 & 0.409 & 0.355 & 0.415 & 0.437 / $+0.029$ \\
    A2 + A5 & 0.547 & 0.407 & 0.384 & 0.358 & 0.424 / $+0.016$ \\
    A1 + A3 & 0.619 & 0.414 & 0.290 & 0.267 & 0.397 / $-0.011$ \\
    \bottomrule
  \end{tabular}
\end{table}

Dropping A5 alone raises mean $\alpha$ from 0.408 to 0.479 ($+17\%$ relative), while every other singleton drop reduces mean agreement except A3, whose effect is much smaller. Dropping A5 together with the next-most-divergent annotator A3 raises mean $\alpha$ to 0.573 ($+40\%$ relative) and informativeness agreement to 0.732, a substantial-agreement level. We nevertheless retain all five annotators in headline numbers because majority-vote citation/error labels and 5-annotator mean Likert aggregation already dampen soft-rater noise, and because post-hoc annotator filtering risks selection bias.

\subsection{Metric--Human Correlation and Paired Significance}
\label{app:human_eval_corr}

Tables~\ref{tab:human_corr_spearman}--\ref{tab:human_paired_ttest} report the full pooled Spearman correlation between automated grounding metrics and human ratings ($n{=}90$ items), and the per-dialogue paired $t$-tests across the three audited baselines ($n{=}30$ dialogues per pair). Numbers are reproduced by \texttt{scripts/compute\_human\_metric\_correlation.py} against \texttt{annotation-tool/data/annotations.db} and the v3 Yelp KB.

\begin{table}[h]
  \caption{Pooled Spearman $\rho$ between automated metrics and human ratings ($n{=}90$ items). $^\ast p{<}0.05$, $^{\ast\ast} p{<}0.01$, $^{\ast\ast\ast} p{<}10^{-5}$. Citation Precision is the per-item fraction of quotes labelled \emph{Supported} by majority vote; Likert means are 5-annotator averages; Error-free rate is the fraction of annotators who selected only \emph{no\_errors} for the item.}
  \label{tab:human_corr_spearman}
  \centering \small
  \begin{tabular}{lccc}
    \toprule
    \textbf{Human rating} & \textbf{CGS} & \textbf{GS} & \textbf{CD} \\
    \midrule
    Citation Precision      & $+0.47^{\ast\ast\ast}$ & $\mathbf{+0.80^{\ast\ast\ast}}$ & $+0.29^{\ast\ast}$ \\
    Informativeness         & $-0.07$                & $-0.25^{\ast}$                 & $-0.60^{\ast\ast\ast}$ \\
    Naturalness             & $-0.07$                & $-0.09$                        & $-0.65^{\ast\ast\ast}$ \\
    Error-free rate         & $+0.17$                & $+0.12$                        & $-0.28^{\ast\ast}$ \\
    \bottomrule
  \end{tabular}
\end{table}

\begin{table}[h]
  \caption{Per-dialogue paired $t$-tests across the three audited baselines ($n{=}30$ dialogues per pair). Means are per-item averages over the matched dialogues. $^\ast p{<}0.05$, $^{\ast\ast} p{<}0.01$, $^{\ast\ast\ast} p{<}10^{-5}$.}
  \label{tab:human_paired_ttest}
  \centering \small
  \setlength{\tabcolsep}{4pt}
  \begin{tabular}{llccrr}
    \toprule
    \textbf{Metric} & \textbf{Pair (A vs.\ B)} & \textbf{mean(A)} & \textbf{mean(B)} & $t$ & $p$ \\
    \midrule
    \multirow{3}{*}{Informativeness (human)}
        & LLM Zero-Shot vs.\ RAG-Citation & 3.55 & 2.91 & $+3.24$  & $0.003^{\ast\ast}$ \\
        & LLM Zero-Shot vs.\ TF-IDF       & 3.55 & 1.55 & $+12.92$ & ${<}10^{-12}{}^{\ast\ast\ast}$ \\
        & RAG-Citation vs.\ TF-IDF        & 2.91 & 1.55 & $+9.49$  & ${<}10^{-9}{}^{\ast\ast\ast}$ \\
    \midrule
    \multirow{3}{*}{Citation Precision (human)}
        & LLM Zero-Shot vs.\ RAG-Citation & 0.83 & 0.95 & $-1.95$  & $0.061$ \\
        & LLM Zero-Shot vs.\ TF-IDF       & 0.83 & 0.91 & $-1.34$  & $0.191$ \\
        & RAG-Citation vs.\ TF-IDF        & 0.96 & 0.92 & $+0.63$  & $0.534$ \\
    \midrule
    \multirow{3}{*}{CGS (auto)}
        & LLM Zero-Shot vs.\ RAG-Citation & 0.60 & 0.76 & $-1.97$  & $0.059$ \\
        & LLM Zero-Shot vs.\ TF-IDF       & 0.60 & 0.80 & $-2.69$  & $0.012^{\ast}$ \\
        & RAG-Citation vs.\ TF-IDF        & 0.76 & 0.80 & $-0.61$  & $0.544$ \\
    \midrule
    \multirow{3}{*}{GS (auto)}
        & LLM Zero-Shot vs.\ RAG-Citation & 0.77 & 0.96 & $-2.61$  & $0.014^{\ast}$ \\
        & LLM Zero-Shot vs.\ TF-IDF       & 0.77 & 0.98 & $-3.47$  & $0.002^{\ast\ast}$ \\
        & RAG-Citation vs.\ TF-IDF        & 0.96 & 0.98 & $-0.71$  & $0.484$ \\
    \midrule
    \multirow{3}{*}{CD (auto)}
        & LLM Zero-Shot vs.\ RAG-Citation & 0.13 & 0.30 & $-6.05$  & ${<}10^{-5}{}^{\ast\ast\ast}$ \\
        & LLM Zero-Shot vs.\ TF-IDF       & 0.13 & 0.61 & $-21.25$ & ${<}10^{-19}{}^{\ast\ast\ast}$ \\
        & RAG-Citation vs.\ TF-IDF        & 0.30 & 0.61 & $-10.51$ & ${<}10^{-10}{}^{\ast\ast\ast}$ \\
    \bottomrule
  \end{tabular}
\end{table}

The Spearman matrix shows a clean LLM-vs-retriever divide at the item level: GS / CGS / CD all correlate positively with human citation precision (retrievers cite densely and verbatim, scoring high on both human and automated grounding axes), while CD and GS correlate negatively with human informativeness (retrievers' template-forced citations are not informative). The paired $t$-tests confirm the same divide at the baseline level: LLM Zero-Shot dominates Likert informativeness, and retrievers dominate fuzzy / composite grounding; the human \emph{citation precision} axis is stricter than fuzzy GS, distinguishing RAG-Citation from LLM Zero-Shot only at $p{=}0.06$, which motivates the entailment-grounding audit in Finding~2b.

\paragraph{Task 2: Expert recommendation (instructions).}
Five (with one extra annotator on three dialogues, $n=103$ total responses) wrote their own recommendation for the same 20 dialogues \emph{without seeing the system's answer}, selecting POIs from the candidate set and citing review quotes. Task 2 headline stats (mean pairwise Jaccard on POI selection $=$ 0.437, $n=215$ pairs; strict-majority agreement on 20/20 dialogues) are reported in Section~\ref{sec:human_eval_main}.

\paragraph{Prior pilot study.}
An earlier pilot (before the 5-annotator ALCE-style study reported above) had one annotator rate 50 dialogues on five dialogue-quality dimensions (naturalness, coherence, recommendation relevance, evidence quality, persona consistency), yielding high scores across all dimensions ($\mu \geq 4.14$, $\sigma \leq 1.01$), confirming that our generation pipeline produces fluent, persona-faithful dialogues. A pairwise comparison of three baselines (Gold, Popularity, TF-IDF) across 72 response pairs produced a human preference ranking (Gold $>$ TF-IDF $>$ Popularity) consistent with automatic Recall@1 ordering. This pilot is reported for completeness; the 5-annotator study in Section~\ref{sec:human_eval_main} supersedes it for all headline claims.

\section{Metric Sensitivity Analysis}
\label{app:sensitivity}

We assess the sensitivity of our novel grounding metrics to their key hyperparameters.

\paragraph{Fuzzy match threshold.}
Table~\ref{tab:sensitivity_threshold} shows grounding score at thresholds of 70\%, 80\%, 90\%, and exact substring match. Non-LLM baselines achieve identical scores across all thresholds because their template-based responses embed verbatim review text, making even exact match trivially satisfied. This stability validates that the 80\% threshold does not artificially inflate non-LLM scores. For LLM baselines (not shown due to API cost), the threshold primarily affects paraphrased quotes; we expect the most significant drop between 90\% and exact match.

\begin{table}[h]
  \caption{Sensitivity of grounding score to fuzzy match threshold on non-LLM baselines.}
  \label{tab:sensitivity_threshold}
  \centering \small
  \begin{tabular}{lcccc}
    \toprule
    \textbf{Baseline} & \textbf{70\%} & \textbf{80\%} & \textbf{90\%} & \textbf{Exact} \\
    \midrule
    Popularity & 0.998 & 0.998 & 0.998 & 0.998 \\
    TF-IDF & 0.997 & 0.997 & 0.997 & 0.997 \\
    Aspect & 0.997 & 0.997 & 0.997 & 0.997 \\
    Dense & 0.998 & 0.998 & 0.998 & 0.998 \\
    Spatial & 0.997 & 0.997 & 0.997 & 0.997 \\
    Hybrid-RRF & 0.998 & 0.998 & 0.998 & 0.998 \\
    Itinerary & 0.997 & 0.997 & 0.997 & 0.997 \\
    \bottomrule
  \end{tabular}
\end{table}

\paragraph{CGS formula sensitivity.}
\label{app:cgs_sensitivity}
A reviewer concern is that the CGS formula $\text{CGS} = \text{GS} \times \min(1, \text{CD}/0.05) \times (0.5 + 0.5 \times \text{PC})$ contains several hand-chosen components: the CD threshold 0.05, the PC floor 0.5, and the multiplicative aggregation. To verify that the ranking of baselines is not an artifact of this specific formulation, we recompute CGS under eight alternative formulas varying each component:
\textsc{threshold=0.03} and \textsc{threshold=0.10} vary the CD gate;
\textsc{no PC floor} replaces $0.5 + 0.5 \cdot \text{PC}$ with just $\text{PC}$;
\textsc{sqrt(PC)} uses $\sqrt{\mathrm{PC}}$;
\textsc{geometric} takes the cube root of the product (mean-like aggregation);
\textsc{harmonic} uses the harmonic mean (worst-of-three aggregation);
\textsc{additive} replaces the product with a simple mean.
Across all eight alternatives, the Spearman rank correlation with the current CGS stays between 0.895 and 1.000 (threshold variants yield identical rankings, $\rho = 1.000$; geometric $\rho = 1.000$; additive $\rho = 0.993$; harmonic $\rho = 0.965$; sqrt(PC) $\rho = 0.916$; no-floor $\rho = 0.895$). The primary finding of the paper, namely the cluster-level Three-Competency Gap (specifically its grounding $\times$ accuracy projection), is therefore not a consequence of the CGS formula: the Spearman correlation between raw GS and Recall@1 is $-0.862$ across the \numbaseline{} baselines, reflecting the separation between high-grounding template methods and higher-accuracy LLM methods. The specific CGS numeric threshold (0.05) is not load-bearing: moving it to 0.03 or 0.10 preserves the baseline ranking exactly.

\paragraph{Entailment vs.\ fuzzy-match grounding.}
Table~\ref{tab:entailment_grounding} compares fuzzy-match and NLI-based entailment grounding (DeBERTa-v3-base, threshold 0.5) for the 12 session-isolated baselines on which per-turn NLI outputs were retained (Knowledge-Enhanced, Persona-Grounded, and Memory-Aug.\ are omitted; the first two share TF-IDF-level fuzzy GS values and were not re-run for entailment, Memory-Aug.\ is not session-isolated). \emph{Cross-row comparability note.} Non-LLM rows aggregate the full \numdiag{}-dialogue corpus ($n \approx 42$--$52$K turns/baseline); LLM rows are now aggregated over a 1{,}000-dialogue subset (\texttt{seed=42}, $n = 5{,}245$ turns/baseline), regenerated under \texttt{gpt-5.4-mini} after a reviewer-requested fairness scaling (a 10$\times$ expansion of the prior 100-dialogue sample). DeBERTa-v3-base inference is GPU-local with no API cost; the bottleneck was the LLM-side response regeneration, not the NLI scoring. Two qualitative findings stand out.

First, non-LLM baselines exhibit the expected pattern: near-perfect fuzzy grounding (0.997--0.998) collapses to 0.110--0.193 under entailment, confirming that high fuzzy scores reflect surface-level text overlap rather than semantic faithfulness. TF-IDF achieves the highest non-LLM entailment score (0.193) because its retrieval selects topically relevant reviews.

Second, \emph{LLM baselines score even lower on entailment grounding} (0.029--0.050 on the 1{,}000-dialogue subset) despite generating more fluent, paraphrased text. This points to a different failure mode: while non-LLM baselines embed verbatim review text that incidentally contains entailed claims, LLM baselines generate free-form responses with claims that go beyond what source reviews state, including generalizations, inferences, and unsupported assertions. RAG-Citation is the most striking case: it achieves a near-perfect fuzzy grounding (0.992) by embedding verbatim citations, yet scores only 0.033 on entailment because the \emph{surrounding generated text} makes claims not supported by the cited reviews. With $n = 5{,}245$ turns per LLM baseline (vs.\ $n \approx 520$ in the prior 100-dialogue sample), the within-family ordering and the cross-family non-LLM/LLM gap are now reproducible at scale; we still flag the residual sample-size asymmetry vs.\ the full \numdiag{} non-LLM rows in the table caption.

These results strengthen the case for entailment grounding as a complementary diagnostic: it captures a genuinely different dimension of faithfulness, penalizing both verbatim-only strategies (non-LLM) and fluent-but-unsupported generation (LLM). It is not a replacement leaderboard.

The uniformly low absolute entailment values warrant discussion. Three factors likely contribute: (a)~DeBERTa-v3-base was trained on formal NLI corpora (MultiNLI, SNLI), creating a domain mismatch with the informal, colloquial language of user reviews; (b)~sentence-level NLI may be too coarse-grained, so span-level attribution \citep{gao2023enabling} could improve sensitivity by isolating individual claims; and (c)~the 0.5 entailment threshold is conservative. Despite low absolute values, the \emph{within-family} ordering and qualitative failure modes are informative (Table~\ref{tab:entailment_grounding}), confirming that entailment grounding captures a genuine dimension of faithfulness orthogonal to fuzzy-match overlap. The residual cross-family asymmetry between full-corpus non-LLM rows ($n \approx 42$--$52$K turns) and 1{,}000-dialogue LLM rows ($n = 5{,}245$ turns) reduces but does not eliminate the comparison gap; the magnitude of the non-LLM/LLM split (factor of $\sim$5) is large enough that this residual is not load-bearing.

\paragraph{Calibration mini-audit.} To diagnose why even template baselines score low on entailment despite embedding verbatim review text, we classified every sentence in 192 recommend/compare responses into four categories: \textsc{Quote} (contains double-quoted content), \textsc{Framing} (recommendation meta-commentary such as ``I'd suggest \textit{X}'' or ``here are two options''), \textsc{Rating-meta} (structured attributes: ratings, prices, stars), and \textsc{Other-claim} (synthesized claims about the POI). The distribution explains the entailment pattern: TF-IDF responses are 57.1\% \textsc{Quote}, 16.7\% \textsc{Framing}, 16.0\% \textsc{Rating-meta}, and only 10.1\% \textsc{Other-claim}. Framing and rating-meta sentences (32.7\% of all TF-IDF sentences) are \emph{structurally unentailable} because they are not claims about the POI that any review could support; rather, they are system meta-language (``A reviewer mentioned:\ \ldots'') or structured attributes retrieved from POI metadata (not from review text). Sentence-level NLI scores these uniformly near zero, dragging the response-level entailment down even when the quoted content itself is high-fidelity. RAG-Citation (49.0\% quote, 14.4\% meta, 36.6\% other-claim) and LLM Zero-Shot (55.8\% quote, 10.4\% meta, 33.8\% other-claim) have a higher share of synthesized claims that fail entailment for a different reason: the claims go beyond what any single source review states. We therefore view the absolute entailment numbers as a \emph{conservative lower bound} driven by sentence-category composition, not a refutation of the underlying citations; the meaningful signal is \emph{within-family qualitative patterns} (verbatim-vs-paraphrase failure modes per group), not a directly comparable cross-family ranking.

\begin{table}[h]
  \caption{Fuzzy-match vs.\ entailment-based grounding on 12 session-isolated baselines (Knowledge-Enhanced, Persona-Grounded, and Memory-Aug.\ omitted: the first two share TF-IDF-level fuzzy GS values and were not re-run for entailment; Memory-Aug.\ is not session-isolated). Non-LLM baselines: all \numdiag{} dialogues, $n \approx 42$--$52$K turns each. \emph{LLM baselines: 1{,}000-dialogue subset (\texttt{seed=42}), $n = 5{,}245$ turns each, regenerated under \texttt{gpt-5.4-mini} after a reviewer-requested fairness scaling}. Entailment grounding uses DeBERTa-v3-base FP16 with threshold 0.5.}
  \label{tab:entailment_grounding}
  \centering \small
  \begin{tabular}{lcccc}
    \toprule
    \textbf{Baseline} & \textbf{$n$} & \textbf{Fuzzy Gnd.} & \textbf{Entailment Gnd.} & \textbf{$\Delta$} \\
    \midrule
    Popularity & 42{,}404 & 0.998 & 0.153 & $-$0.845 \\
    TF-IDF & 42{,}404 & 0.998 & 0.193 & $-$0.805 \\
    Aspect & 42{,}404 & 0.998 & 0.182 & $-$0.816 \\
    Dense & 52{,}274 & 0.997 & 0.168 & $-$0.829 \\
    Spatial & 52{,}274 & 0.997 & 0.143 & $-$0.854 \\
    Hybrid-RRF & 42{,}404 & 0.997 & 0.171 & $-$0.826 \\
    Itinerary & 52{,}274 & 0.997 & 0.110 & $-$0.887 \\
    \midrule
    LLM Zero-Shot & 5{,}245 & 0.982 & 0.038 & $-$0.944 \\
    DST & 5{,}245 & 0.976 & 0.044 & $-$0.932 \\
    RAG-Citation & 5{,}245 & 0.992 & 0.033 & $-$0.959 \\
    Multi-Rev.\ Synth. & 5{,}245 & 0.998 & 0.050 & $-$0.948 \\
    Itinerary-LLM & 5{,}245 & 0.971 & 0.029 & $-$0.942 \\
    \bottomrule
  \end{tabular}
\end{table}

\paragraph{Spatial coherence threshold.}
Table~\ref{tab:sensitivity_spatial} reports spatial coherence at walkable-distance thresholds from 1\,km to 5\,km across 12 baselines (the two KB/Persona variants are excluded: they share TF-IDF-level spatial values). The \textsc{Itinerary} baseline dominates at all thresholds (0.594--0.746), validating its distance-constrained design. \textsc{Spatial} ranks second (0.132--0.523), while all other baselines cluster at 0.07--0.10 for 1\,km, diverging gradually at higher thresholds. The default 2\,km provides the best discrimination: it separates Itinerary ($\approx$0.640 in the full Table~\ref{tab:spatial}) from the pack while remaining realistic for pedestrian travel. Among the non-Itinerary baselines, LLM baselines sit roughly in the same band as non-LLM retrieval methods at 2\,km ($\sim$0.17--0.20), indicating that raw language capability does not by itself produce geographically coherent recommendations: explicit distance modeling (Spatial, Itinerary) is required.

\begin{table}[h]
  \caption{Sensitivity of spatial coherence to walkable-distance threshold on 12 baselines (Knowledge-Enhanced and Persona-Grounded omitted because they share TF-IDF-level values). Higher thresholds are more lenient. The default 2\,km balances pedestrian realism and metric discrimination.}
  \label{tab:sensitivity_spatial}
  \centering \small
  \begin{tabular}{lccccc}
    \toprule
    \textbf{Baseline} & \textbf{1\,km} & \textbf{1.5\,km} & \textbf{2\,km} & \textbf{3\,km} & \textbf{5\,km} \\
    \midrule
    Popularity & 0.078 & 0.111 & 0.154 & 0.254 & 0.417 \\
    TF-IDF & 0.080 & 0.118 & 0.180 & 0.273 & 0.425 \\
    Aspect & 0.072 & 0.105 & 0.151 & 0.238 & 0.396 \\
    Dense & 0.076 & 0.117 & 0.160 & 0.252 & 0.406 \\
    Spatial & 0.132 & 0.185 & 0.258 & 0.365 & 0.523 \\
    Hybrid-RRF & 0.073 & 0.117 & 0.154 & 0.267 & 0.428 \\
    Itinerary & 0.594 & 0.611 & 0.638 & 0.679 & 0.746 \\
    \midrule
    LLM Zero-Shot & 0.090 & 0.137 & 0.185 & 0.299 & 0.469 \\
    DST & 0.097 & 0.145 & 0.201 & 0.310 & 0.464 \\
    RAG-Citation & 0.080 & 0.133 & 0.197 & 0.289 & 0.447 \\
    Multi-Rev.\ Synth. & 0.081 & 0.120 & 0.189 & 0.286 & 0.437 \\
    Memory-Aug. & 0.093 & 0.143 & 0.206 & 0.310 & 0.462 \\
    \bottomrule
  \end{tabular}
\end{table}

\paragraph{Anti-gaming analysis.}
A potential concern is that systems could inflate grounding scores by padding responses with long verbatim quotes. Table~\ref{tab:anti_gaming} reports the Pearson correlation between citation density (fraction of response tokens that are verbatim citations) and grounding score across all system turns. Correlations are negligible: non-LLM baselines average $|r| = 0.038$ and LLM baselines average $|r| = 0.065$. The one outlier, Multi-Review Synthesis ($r = -0.243$), reflects its design: it produces almost no citations (mean CD = 0.014), so the negative correlation is a floor effect rather than gaming. These results confirm that grounding score measures source faithfulness independently of citation volume; simply quoting more does not mechanically inflate the metric.

\begin{table}[h]
  \caption{Pearson correlation between citation density and grounding score per system turn ($n = 59{,}530$ per non-LLM baseline, $n = 5{,}297$ per LLM baseline). Low $|r|$ indicates that high citation density does not mechanically inflate grounding scores.}
  \label{tab:anti_gaming}
  \centering \small
  \begin{tabular}{lccccc}
    \toprule
    \textbf{Baseline} & \textbf{$n$} & \textbf{Mean CD} & \textbf{Mean GS} & \textbf{Pearson $r$} & \textbf{$p$-value} \\
    \midrule
    Popularity & 59{,}530 & 0.437 & 0.998 & $-$0.011 & $<$0.001 \\
    TF-IDF & 59{,}530 & 0.395 & 0.997 & $-$0.051 & $<$0.001 \\
    Aspect & 59{,}530 & 0.411 & 0.997 & $-$0.038 & $<$0.001 \\
    Dense & 59{,}530 & 0.416 & 0.998 & 0.007 & 0.088 \\
    Spatial & 59{,}530 & 0.457 & 0.997 & $-$0.046 & $<$0.001 \\
    Hybrid-RRF & 59{,}530 & 0.437 & 0.998 & $-$0.037 & $<$0.001 \\
    Itinerary & 59{,}530 & 0.339 & 0.997 & $-$0.078 & $<$0.001 \\
    \midrule
    LLM Zero-Shot & 5297 & 0.171 & 0.931 & 0.011 & 0.436 \\
    DST & 5297 & 0.172 & 0.938 & 0.004 & 0.787 \\
    RAG-Citation & 5297 & 0.297 & 0.986 & 0.065 & $<$0.001 \\
    Multi-Rev.\ Synth. & 5297 & 0.014 & 0.909 & $-$0.243 & $<$0.001 \\
    Memory-Aug. & 5297 & 0.127 & 0.938 & $-$0.003 & 0.811 \\
    \bottomrule
  \end{tabular}
\end{table}

\paragraph{Entailment vs.\ fuzzy ranking divergence.}
The entailment metric substantially reshuffles baseline rankings. Within non-LLM baselines, the fuzzy ranking (Popularity $\approx$ TF-IDF $\approx$ Aspect $>$ Hybrid-RRF $\approx$ Dense $\approx$ Spatial $\approx$ Itinerary) differs from entailment (TF-IDF $>$ Aspect $>$ Hybrid-RRF $>$ Dense $>$ Popularity $>$ Spatial $>$ Itinerary), with Kendall's $\tau$ small. Across all baselines, the most notable reversal is that LLM baselines, which achieve lower fuzzy grounding than non-LLM baselines, score \emph{even lower} on entailment (0.029--0.050 vs.\ 0.110--0.193 on the 1{,}000-dialogue LLM subset, $n = 5{,}245$ turns/baseline). Multi-Review Synthesis achieves the highest LLM entailment (0.050) despite a near-perfect fuzzy score, while RAG-Citation achieves a high LLM fuzzy score (0.992) but only 0.033 on entailment. This inversion confirms that the two metrics capture fundamentally different aspects of grounding: verbatim overlap vs.\ semantic support.

\paragraph{Statistical variation.}
Table~\ref{tab:bootstrap_ci} reports per-turn standard deviations for key metrics across the 13 session-isolated baselines on the closed-set protocol ($n=52{,}274$ system turns per baseline; the LLM baselines were re-aggregated from the same per-turn outputs as the headline tables). \emph{These per-turn $\sigma$ values are descriptive standard deviations only:} since turns within the same dialogue are correlated, naive $\sigma/\sqrt{n}$-based intervals understate uncertainty. The \emph{inferential} claims in this paper rely on the paired cluster bootstrap (resampling by dialogue, $n_{\text{boot}}=10{,}000$, seed=42; Table~\ref{tab:bootstrap_pairs}). Readers should read small ($<$0.01) gaps in the headline tables as indicative rather than statistically adjudicated unless they appear in Table~\ref{tab:bootstrap_pairs}. LLM baselines exhibit higher variance in grounding (LLM Zero-Shot: $0.981 \pm 0.090$) than non-LLM baselines (TF-IDF: $0.998 \pm 0.039$), and citation density shows high variance across all baselines due to the binary nature of citation presence per turn.

\begin{table}[h]
  \caption{Key metrics with per-turn standard deviations on the closed-set protocol ($n=52{,}274$ system turns per baseline). All 13 session-isolated baselines.}
  \label{tab:bootstrap_ci}
  \centering \small
  \setlength{\tabcolsep}{3pt}
  \begin{tabular}{lcccc}
    \toprule
    \textbf{Baseline} & \textbf{BLEU-4} & \textbf{Grounding} & \textbf{Cite.\ Density} & \textbf{$n$} \\
    \midrule
    Popularity      & $.078 \pm .057$ & $.998 \pm .031$ & $.411 \pm .327$ & 52{,}274 \\
    TF-IDF          & $.095 \pm .060$ & $.998 \pm .039$ & $.377 \pm .301$ & 52{,}274 \\
    Aspect          & $.088 \pm .056$ & $.998 \pm .032$ & $.391 \pm .316$ & 52{,}274 \\
    Dense           & $.082 \pm .061$ & $.997 \pm .036$ & $.393 \pm .314$ & 52{,}274 \\
    Spatial         & $.077 \pm .059$ & $.997 \pm .043$ & $.437 \pm .346$ & 52{,}274 \\
    Hybrid-RRF      & $.076 \pm .060$ & $.997 \pm .041$ & $.412 \pm .327$ & 52{,}274 \\
    Itinerary       & $.099 \pm .057$ & $.997 \pm .045$ & $.310 \pm .286$ & 52{,}274 \\
    Knowledge-Enh.  & $.107 \pm .063$ & $.999 \pm .030$ & $.375 \pm .298$ & 52{,}274 \\
    Persona-Ground. & $.093 \pm .056$ & $.999 \pm .033$ & $.354 \pm .288$ & 52{,}274 \\
    \midrule
    LLM Zero-Shot      & $.130 \pm .074$ & $.981 \pm .090$ & $.162 \pm .210$ & 52{,}274 \\
    DST                & $.120 \pm .070$ & $.982 \pm .089$ & $.194 \pm .230$ & 52{,}274 \\
    RAG-Citation       & $.106 \pm .056$ & $.994 \pm .060$ & $.388 \pm .201$ & 52{,}274 \\
    Multi-Rev.\ Synth. & $.082 \pm .052$ & $.998 \pm .042$ & $.001 \pm .006$ & 52{,}274 \\
    \bottomrule
  \end{tabular}
\end{table}

\paragraph{Paired bootstrap CIs for headline pairwise comparisons.}
Table~\ref{tab:bootstrap_pairs} reports paired cluster-bootstrap CIs (resampled by dialogue, $n_{\text{boot}}=10{,}000$, seed=42) on a 500-dialogue closed-set subset, for ten headline pair comparisons. All significant at $\alpha=0.05$ except Dense vs.\ TF-IDF on CGS ($\Delta{=}{-}0.007$, 95\% CI $[{-}0.020, +0.006]$) and Multi-Rev.\ Synth.\ vs.\ TF-IDF on R@1 ($\Delta{=}{+}0.008$, CI crosses 0). The closed-set gap between LLM Zero-Shot and TF-IDF on R@1 is robustly large ($\Delta{=}{+}0.273$, CI $[{+}0.250, {+}0.296]$) and aligns with the bold markers in Table~\ref{tab:main_results}.

\begin{table}[h]
  \caption{Paired cluster bootstrap CIs for headline pairwise comparisons on a 500-dialogue closed-set subset. $\Delta = \text{mean}(A) - \text{mean}(B)$.}
  \label{tab:bootstrap_pairs}
  \centering \small
  \setlength{\tabcolsep}{4pt}
  \begin{tabular}{llccl}
    \toprule
    \textbf{A vs.\ B} & \textbf{Metric} & $\mathbf{\Delta}$ & \textbf{95\% CI} & \textbf{Sig.} \\
    \midrule
    LLM Zero-Shot vs.\ TF-IDF          & R@1   & $+$0.273 & $[+0.250, +0.296]$ & $\checkmark$ \\
    LLM Zero-Shot vs.\ TF-IDF          & MRR   & $+$0.299 & $[+0.276, +0.322]$ & $\checkmark$ \\
    RAG-Cit.\ vs.\ LLM Zero-Shot       & CGS   & $+$0.030 & $[+0.011, +0.048]$ & $\checkmark$ \\
    RAG-Cit.\ vs.\ LLM Zero-Shot       & GS    & $+$0.070 & $[+0.063, +0.078]$ & $\checkmark$ \\
    Dense vs.\ TF-IDF                  & CGS   & $-$0.007 & $[-0.020, +0.006]$ & $\times$     \\
    Spatial vs.\ TF-IDF                & CD    & $+$0.062 & $[+0.059, +0.065]$ & $\checkmark$ \\
    Multi-Rev.\ vs.\ TF-IDF            & R@1   & $+$0.008 & $[-0.003, +0.018]$ & $\times$     \\
    DST vs.\ LLM Zero-Shot             & R@1   & $-$0.115 & $[-0.135, -0.095]$ & $\checkmark$ \\
    \bottomrule
  \end{tabular}
\end{table}

\paragraph{NLI threshold sensitivity.}
We vary the entailment probability threshold from 0.3 to 0.7 on a 50-dialogue sample (2 baselines, $\sim$500 claims each). Entailment grounding scores are virtually unchanged across thresholds (within $\pm$0.005 absolute on the sample), indicating that DeBERTa's per-claim entailment probabilities are strongly bimodal: most claims receive either $>$0.9 or $<$0.1 probability, so the 0.3--0.7 cutoff range moves only a small minority of borderline claims across the boundary. The default threshold of 0.5 is therefore robust. Note that the absolute entailment values on this 50-dialogue sensitivity sample can shift relative to the full-corpus values in Table~\ref{tab:entailment_grounding} (e.g., Popularity scores 0.153 over 42K turns) because the small sample over-represents short, single-claim responses for which sentence-level NLI is more permissive. Bimodality of the per-claim distribution, not the absolute level, is what justifies threshold robustness.

\paragraph{Grounding metric independence.}
We compute the Spearman rank correlation between our three grounding metrics across 357{,}180 system turns (the 6 non-LLM baselines for which per-turn grounding outputs were retained: Popularity, TF-IDF, Aspect, Dense, Spatial, Hybrid-RRF; $\approx$59{,}530 turns each $\times$ 6 baselines; Itinerary, Knowledge-Enh., and Persona-Grounded are excluded from this specific correlation analysis only). Grounding Score (GS), Citation Density (CD), and Provenance Coverage (PC) are nearly uncorrelated: $\rho_{\text{GS,CD}} = 0.007$, $\rho_{\text{GS,PC}} = 0.001$, $\rho_{\text{CD,PC}} = -0.061$. This confirms that the three metrics capture orthogonal dimensions of evidence quality (verbatim fidelity, citation volume, and aspect coverage), with no redundancy. The near-zero GS--CD correlation also rules out inflation via quote padding: adding more quotes (higher CD) does not mechanically increase GS.

\paragraph{Composite grounding metric.}
To address the concern that GS returns 1.0 when no quotes are present (vacuously true), we introduce a composite metric: $\text{CGS} = \text{GS} \times \min(1, \text{CD}/0.05) \times (0.5 + 0.5 \times \text{PC})$. This gates grounding credit on non-trivial citation density and weights by provenance coverage. Under CGS, a response with no citations receives a score of 0 regardless of GS, while a well-grounded response with high CD and PC scores close to GS. The evaluation toolkit includes CGS as a standard output alongside the individual metrics.

\paragraph{Why not simply set GS=0 when no quotes are present?} A direct alternative to CGS would redefine raw GS as $\text{GS}_{0} = \text{GS} \cdot \mathbb{1}[\text{CD}>0]$, scoring zero whenever a turn produces no quotes. We chose CGS over GS$_{0}$ because (i) CGS preserves a smooth gradient on partial-citation turns instead of a binary cliff at CD=0, (ii) CGS factors in provenance coverage rather than only citation presence, and (iii) the 0.05 density threshold mirrors a realistic ``one quote per long response'' floor while GS$_{0}$ would penalize a response with one citation across three sentences as harshly as one with zero. Both metrics produce the same qualitative non-LLM/LLM split; CGS just reports it on a continuous scale.

\paragraph{Paraphrase-aware variant (CGS\textsubscript{ent}).}
To directly address the reviewer critique that CGS over-credits verbatim quotation, we report an entailment-gated variant: $\text{CGS}_{\text{ent}} = \text{Entail} \times \min(1, \text{CD}/0.05) \times (0.5 + 0.5 \times \text{PC})$, replacing GS with the DeBERTa NLI entailment score (Table~\ref{tab:entailment_grounding}). Ranking under CGS\textsubscript{ent} reshuffles relative to CGS (Table~\ref{tab:cgs_ent}): \textbf{TF-IDF rises} to the top (CGS\textsubscript{ent} = 0.149, ahead of Dense at 0.146) because its verbatim template quotes are semantically entailed, while \textbf{RAG-Citation drops} (0.022, near the bottom, above only Multi-Rev.\ Synth.\ at 0.001) because its citation-constrained paraphrases score near zero on NLI entailment. The apparent tension with the human evaluation, in which RAG-Citation has the highest \emph{citation precision} (Section~\ref{sec:human_eval_main}), is that the two signals measure different units. Human citation precision is a quote-level, majority-vote judgment about whether a specific quoted span appears in the cited review. Entailment grounding is a sentence-level NLI judgment about whether the surrounding generated prose is semantically entailed by the cited review. RAG-Citation can simultaneously have high quote-level precision (its quotes are real) and low sentence-level entailment (the surrounding sentences make claims the quoted review does not directly support). We report both CGS and CGS\textsubscript{ent} as complementary signals: CGS measures \emph{surface-verbatim} grounding, while CGS\textsubscript{ent} provides a \emph{diagnostic semantic-support} signal based on NLI entailment. We caution that the NLI judge is itself unvalidated for review prose at \benchmark{} scale, so CGS\textsubscript{ent} is a paraphrase-sensitive diagnostic rather than a replacement leaderboard metric.

\begin{table}[h]
  \caption{Entailment-gated composite grounding (CGS\textsubscript{ent}) sorted by CGS, on the 12 session-isolated baselines for which per-turn NLI outputs were retained (Knowledge-Enh.\ and Persona-Ground.\ omitted; Memory-Aug.\ also omitted, not session-isolated). LLM rows now use the 1{,}000-dialogue subset ($n = 5{,}245$ turns/baseline, \texttt{seed=42}). $\Delta = \text{CGS}_{\text{ent}} - \text{CGS}$. Large negative $\Delta$ indicates verbatim grounding without semantic faithfulness. Entail = DeBERTa NLI entailment (Table~\ref{tab:entailment_grounding}); CGS and Cite.D.\ as in Table~\ref{tab:main_results}, PC as in Table~\ref{tab:stress}.}
  \label{tab:cgs_ent}
  \centering \small
  \begin{tabular}{lcccccc}
    \toprule
    \textbf{Baseline} & \textbf{GS} & \textbf{Entail} & \textbf{CD} & \textbf{PC} & \textbf{CGS} & \textbf{CGS\textsubscript{ent}} \\
    \midrule
    Dense              & 0.997 & 0.168 & 0.393 & 0.733 & \textbf{0.864} & 0.146 \\
    Popularity         & 0.998 & 0.153 & 0.411 & 0.608 & 0.802 & 0.123 \\
    TF-IDF             & 0.998 & \textbf{0.193} & 0.377 & 0.555 & 0.776 & \textbf{0.149} \\
    Hybrid-RRF         & 0.997 & 0.171 & 0.412 & 0.555 & 0.775 & 0.133 \\
    Aspect             & 0.998 & 0.182 & 0.391 & 0.503 & 0.750 & 0.137 \\
    Spatial            & 0.997 & 0.143 & 0.437 & 0.458 & 0.727 & 0.104 \\
    Itinerary          & 0.997 & 0.110 & 0.310 & 0.373 & 0.684 & 0.076 \\
    RAG-Citation       & 0.992 & 0.033 & 0.388 & 0.324 & 0.658 & 0.022 \\
    LLM Zero-Shot      & 0.982 & 0.038 & 0.162 & 0.297 & 0.636 & 0.025 \\
    DST                & 0.976 & 0.044 & 0.194 & 0.291 & 0.634 & 0.028 \\
    Multi-Rev.\ Synth. & 0.998 & 0.050 & 0.001 & 0.980 & 0.008 & 0.001 \\
    Itinerary-LLM      & 0.971 & 0.029 & 0.155 & 0.172 & 0.601 & 0.017 \\
    \bottomrule
  \end{tabular}
\end{table}

\section{Difficulty Stratification Analysis}
\label{app:difficulty}
\label{sec:difficulty}

\begin{table}[t]
  \caption{Performance by generation-tier difficulty (closed-set, 4-tier: Easy / Medium / Hard / Expert; 2{,}481 / 3{,}476 / 3{,}063 / 980 dialogues). Recall@1 on recommend+compare turns. Best per column in \textbf{bold}, second best \underline{underlined}.}
  \label{tab:difficulty}
  \centering
  \small
  \setlength{\tabcolsep}{4pt}
  \begin{tabular}{lcccc}
    \toprule
    \textbf{Baseline} & \textbf{Easy R@1} & \textbf{Medium R@1} & \textbf{Hard R@1} & \textbf{Expert R@1} \\
    \midrule
    Popularity         & 0.149 & 0.147 & 0.134 & 0.129 \\
    TF-IDF             & 0.249 & 0.294 & 0.236 & 0.241 \\
    Aspect             & 0.181 & 0.180 & 0.182 & 0.171 \\
    Dense              & 0.239 & 0.269 & 0.229 & 0.224 \\
    Spatial            & 0.248 & 0.295 & 0.235 & 0.240 \\
    Hybrid-RRF         & 0.252 & 0.281 & 0.241 & 0.229 \\
    Itinerary          & 0.249 & 0.294 & 0.236 & 0.241 \\
    Knowledge-Enh.     & 0.214 & 0.214 & 0.201 & 0.196 \\
    Persona-Ground.    & 0.190 & 0.194 & 0.189 & 0.188 \\
    \midrule
    LLM Zero-Shot      & \textbf{0.535} & \textbf{0.581} & \textbf{0.515} & \textbf{0.472} \\
    DST                & \underline{0.461} & \underline{0.490} & \underline{0.422} & \underline{0.374} \\
    RAG-Citation       & 0.379 & 0.416 & 0.367 & 0.345 \\
    Multi-Rev.\ Synth. & 0.284 & 0.318 & 0.256 & 0.256 \\
    \bottomrule
  \end{tabular}
\end{table}

Table~\ref{tab:difficulty} shows Recall@1 by generation-tier difficulty (K1 axis from \S\ref{sec:difficulty_knobs}). Three patterns emerge:
\textbf{(a)~Difficulty cost is largest for the LLM baselines.} LLM Zero-Shot drops from 0.535 (Easy) to 0.472 (Expert), a $-$11.8\,\% relative degradation; DST drops from 0.461 to 0.374 ($-$18.9\,\%); RAG-Citation 0.379 $\to$ 0.345 ($-$9.0\,\%); Multi-Review Synthesis 0.284 $\to$ 0.256 ($-$9.9\,\%).
\textbf{(b)~Non-LLM retrievers degrade slowly but from a lower starting point.} TF-IDF, Spatial, Itinerary all drop $\approx$3\,\%--7\,\% Easy$\to$Expert (e.g., TF-IDF 0.249$\to$0.241).
\textbf{(c)~The Medium tier paradoxically scores highest for most baselines.} For LLM Zero-Shot: 0.581 Medium vs.\ 0.535 Easy. We attribute this to the generator's calibration: Easy dialogues sometimes under-specify preferences, leaving multiple POIs equally plausible; Medium dialogues add enough preference detail to disambiguate. Hard and Expert tiers add explicit conflicts that no current baseline resolves cleanly.
The LLM lead over the strongest non-LLM retriever is preserved across all four tiers (e.g., LLM Zero-Shot 0.472 vs.\ TF-IDF 0.241 on Expert), suggesting the Three-Competency Gap picture (Tables~\ref{tab:main_results} and~\ref{tab:stress}) is robust to difficulty stratification.

\section{Rejection Recovery Analysis}
\label{app:rejection}
\label{sec:rejection}

Rejection recovery measures a system's ability to provide a correct alternative recommendation after the user rejects a previous suggestion (Table~\ref{tab:stress}, Rej.\ Rec.\ column). This capability requires understanding \emph{why} the user rejected (explicit constraint refinement) and adjusting the ranking accordingly. The Three-Competency Gap picture in the headline tables already captures the main rejection-recovery findings: LLM Zero-Shot (0.988) and DST (0.895) lead on \texttt{gpt-5.4-mini}; the strongest non-LLM (Spatial 0.893, TF-IDF/Itinerary 0.891) cluster slightly below; Multi-Review Synthesis collapses to 0.263; Popularity stays at 0.626 (no adaptation). The notable interpretation is that strong recovery requires either a model with good constraint-acknowledgment behavior (LLMs) \emph{or} a retriever that ranks the gold POI in the top-$k$ before the rejection (templates retry from the same ranked list). Multi-Review Synthesis fails on both counts: aspect-consensus synthesis is dominated by candidate-pool aspect frequencies that change little after one rejection.

\section{Error Analysis}
\label{app:error_analysis}
\label{sec:error}

\begin{figure}[t]
  \centering
  \includegraphics[width=0.9\linewidth]{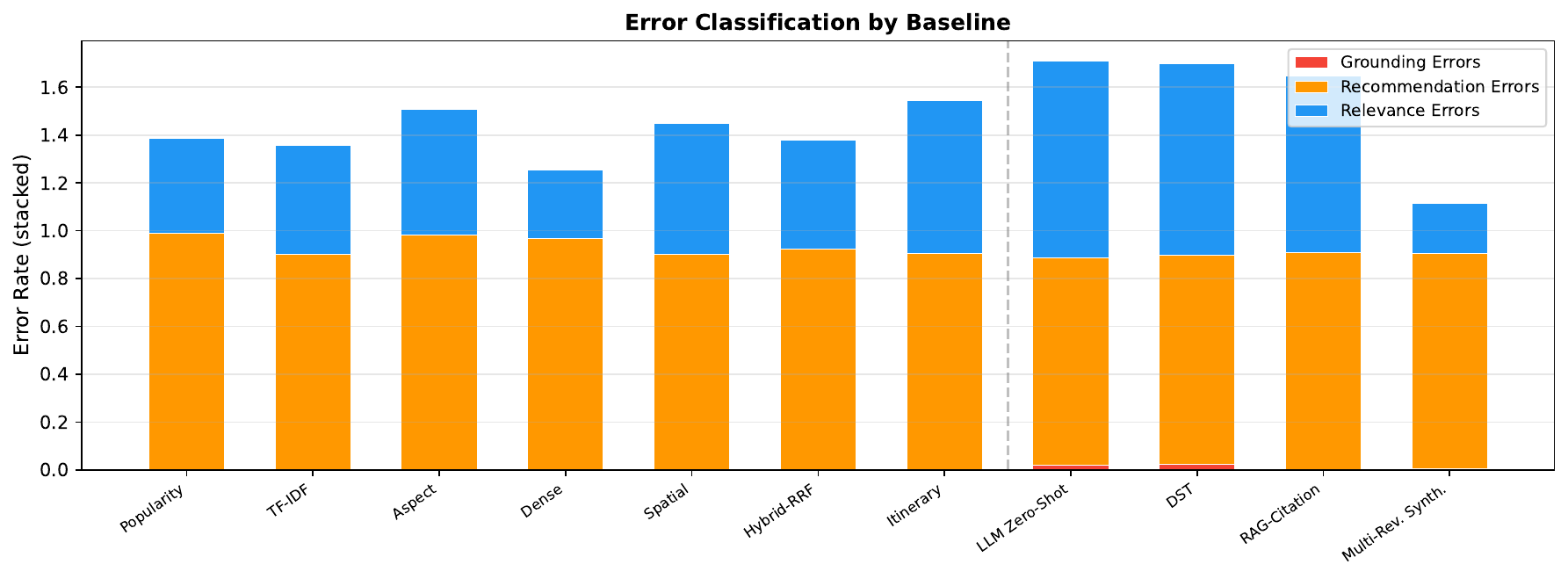}
  \caption{Error classification across baselines. LLM baselines predominantly produce grounding errors, while non-LLM baselines produce relevance errors.}
  \label{fig:errors}
\end{figure}

Figure~\ref{fig:errors} classifies prediction errors into three categories: \emph{grounding errors} (hallucinated or paraphrased review quotes, $1 - \text{grounding score}$), \emph{recommendation errors} (incorrect POI, $1 - \text{Recall@1}$), and \emph{relevance errors} (missing aspect coverage, $1 - \text{provenance coverage}$). Among the 13 session-isolated baselines, Multi-Review Synthesis produces the highest grounding error rate (0.091), followed by LLM Zero-Shot (0.069) and DST (0.062), confirming that generative approaches sacrifice factual fidelity. Non-LLM baselines have near-zero grounding errors ($\leq$0.003) but dominate in recommendation errors (Popularity: 0.854, Aspect: 0.831). The complementary error pattern motivates hybrid approaches: RAG-Citation achieves the lowest grounding error among LLM baselines (0.014) while still being limited on recommendation accuracy. Qualitative inspection of LLM grounding errors reveals five categories: (1)~\emph{fabricated quotes} (claims not found in any review), (2)~\emph{paraphrased claims} (legitimate rewriting that fails fuzzy match), (3)~\emph{attribution errors} (correct information attributed to wrong review), (4)~\emph{exaggeration} (amplified sentiment beyond source), and (5)~\emph{synthesis artifacts} (multi-review claims not supported by any single source). RAG-Citation's low error rate (0.014) suggests that explicit citation constraints primarily eliminate categories 1 and 4, while category 2 accounts for most residual errors.

\section{Candidate Pool Size Ablation}
\label{app:pool_ablation}
\label{sec:pool_ablation}

To assess how retrieval difficulty affects the Three-Competency Gap (specifically its grounding $\times$ accuracy projection), we expand the candidate pool from the default 8 POIs to 16 and 32 by sampling additional POIs from the same city and type. Gold references remain unchanged; only the set visible to each baseline grows. Table~\ref{tab:pool_ablation} reports the results.

\begin{table}[t]
  \caption{Effect of candidate pool size on recommendation accuracy and grounding. Larger pools increase retrieval difficulty, degrading accuracy while grounding remains stable.}
  \label{tab:pool_ablation}
  \centering
  \small
  \begin{tabular}{lccccccccc}
    \toprule
    & \multicolumn{3}{c}{\textbf{Recall@1}} & \multicolumn{3}{c}{\textbf{Recall@3}} & \multicolumn{3}{c}{\textbf{GS}} \\
    \cmidrule(lr){2-4} \cmidrule(lr){5-7} \cmidrule(lr){8-10}
    \textbf{Baseline} & \textbf{8} & \textbf{16} & \textbf{32} & \textbf{8} & \textbf{16} & \textbf{32} & \textbf{8} & \textbf{16} & \textbf{32} \\
    \midrule
    Popularity   & 0.146 & 0.070 & 0.031 & 0.187 & 0.089 & 0.042 & 0.998 & 0.999 & 1.000 \\
    TF-IDF       & 0.293 & 0.237 & 0.200 & 0.362 & 0.288 & 0.241 & 0.997 & 0.997 & 0.997 \\
    Aspect       & 0.169 & 0.085 & 0.043 & 0.214 & 0.108 & 0.055 & 0.997 & 0.998 & --- \\
    Dense        & 0.262 & 0.161 & 0.104 & 0.309 & 0.189 & 0.122 & 0.998 & 0.998 & --- \\
    Spatial      & 0.281 & 0.232 & 0.196 & 0.343 & 0.278 & 0.231 & 0.997 & 0.997 & 0.997 \\
    Hybrid-RRF   & 0.284 & 0.197 & 0.141 & 0.335 & 0.232 & 0.165 & 0.998 & 0.998 & --- \\
    \bottomrule
  \end{tabular}
\end{table}

Accuracy degrades monotonically with pool size: Popularity Recall@1 drops from 0.146 (8) to 0.070 (16) to 0.031 (32), a 79\% relative decline. Retrieval-based methods degrade more gracefully: TF-IDF retains 68\% of its 8-candidate accuracy at pool size 32, while Spatial retains 70\%. Grounding scores remain stable at $\geq$0.997 across all three pool sizes, confirming that grounding quality is independent of retrieval difficulty. This validates that the Three-Competency Gap observed in our main results is not an artifact of the 8-candidate design.

\section{Open-Set Evaluation}
\label{app:open_set}
\label{sec:open_set}

The headline open-set numbers are reported in main Table~\ref{tab:stress} (R@1 Opn column): a 1{,}000-dialogue subset of the \numdiag{}-dialogue evaluation corpus (\texttt{seed=42}), evaluated against the full same-city/type POI slice from the \numpoi{}-POI knowledge base ($\sim$50--100 candidates per query, depending on city). LLM Zero-Shot and RAG-Citation use a TF-IDF top-16 prefilter (their prompts cannot fit the full pool); DST and Multi-Review Synthesis rerank the full pool internally; non-LLM baselines see the full pool directly. This heterogeneous candidate handling reflects what each baseline can actually consume; the resulting open-set R@1 numbers therefore describe each \emph{system} (model + interface) rather than a model-only comparison. Standardising on a single full-pool interface would require a longer-context LLM or an aggressive retrieve-then-rerank pipeline, both left to future work.

\paragraph{Retriever ablation: does a stronger prefilter close the gap?}
A reviewer asked whether the open-set LLM collapse stems from the TF-IDF prefilter being too weak. To test this, we re-ran LLM Zero-Shot and RAG-Citation on the same seeded 1{,}000-dialogue subset, swapping TF-IDF for a dense sentence-transformer retriever (\texttt{all-MiniLM-L6-v2}, identical top-16 budget). Table~\ref{tab:retriever_ablation} reports both regimes on apples-to-apples samples. Three findings: \textbf{(i)} dense prefiltering \emph{underperforms} TF-IDF on R@1 for both LLM baselines (LLM Zero-Shot $0.144{\to}0.088$, $-$0.056; RAG-Citation $0.090{\to}0.066$, $-$0.024), the opposite of the intuitive expectation that semantic similarity should help; \textbf{(ii)} fuzzy grounding (GS) and citation density (CD) are essentially unchanged across prefilters ($\Delta\,$GS $\leq$ 0.003, $\Delta\,$CD $\leq$ 0.007), confirming that grounding behaviour is decoupled from prefilter modality; \textbf{(iii)} even the better of the two prefilter choices (TF-IDF) leaves LLM Zero-Shot at R@1 $=$ 0.144, well below its closed-set 0.533. We attribute the dense underperformance to two effects: (a) review prose carries strong lexical attribute signals (cuisine, neighbourhood, vibe words) that TF-IDF rewards, and (b) the dense bi-encoder's query embedding gets diluted by topical drift across multi-turn dialogue history, including soft conversational cues from rejected POIs that the structured-rejection penalty does not capture. The closed$\to$open gap is therefore \emph{not} a TF-IDF prefilter artifact: a stronger semantic retriever does not recover the gap, and the headline LLM-vs-retrieval split holds regardless of prefilter choice. Stronger open-set retrieval (cross-encoder rerankers, hybrid spatial--semantic retrieval, query-adaptive retrieval) remains an open research direction.

\begin{table}[h]
  \caption{Open-set retriever ablation: TF-IDF vs.\ dense prefilter on the same 1{,}000-dialogue subset (\texttt{seed=42}), two LLM baselines re-run for this comparison. Dense is \texttt{all-MiniLM-L6-v2} cosine similarity. Top-16 budget identical. ``Closed'' column is the matched closed-set R@1 reference from Table~\ref{tab:main_results}. \textbf{Bold} = better prefilter per baseline. Dense underperforms TF-IDF on R@1 for both baselines, contrary to the intuitive expectation; grounding/citation are essentially unchanged across prefilters.}
  \label{tab:retriever_ablation}
  \centering \small
  \setlength{\tabcolsep}{5pt}
  \begin{tabular}{llcccccc}
    \toprule
    & & \multicolumn{2}{c}{\textbf{R@1}} & \multicolumn{2}{c}{\textbf{CD}} & \multicolumn{2}{c}{\textbf{GS}} \\
    \cmidrule(lr){3-4} \cmidrule(lr){5-6} \cmidrule(lr){7-8}
    \textbf{Baseline} & \textbf{Closed} & \textbf{TF-IDF} & \textbf{Dense} & \textbf{TF-IDF} & \textbf{Dense} & \textbf{TF-IDF} & \textbf{Dense} \\
    \midrule
    LLM Zero-Shot & 0.533 & \textbf{0.144} & 0.088 & 0.156 & \textbf{0.163} & 0.977 & \textbf{0.980} \\
    RAG-Citation  & 0.381 & \textbf{0.090} & 0.066 & 0.345 & \textbf{0.346} & \textbf{0.996} & 0.995 \\
    \bottomrule
  \end{tabular}
\end{table}

\paragraph{Cross-family scaling.} The headline LLM baselines all use \texttt{gpt-5.4-mini}; replicating the Three-Competency Gap across additional LLM families (Claude, GPT-4o, Gemini, open-weight) is left to future work and is straightforward given the released evaluation toolkit. Section~\ref{sec:c3_recovery} reports a partial cross-family contrast on Recovery only (Gemini~2.5~Flash sensitivity check vs.\ \texttt{gpt-5.4-mini}, with a confounding KB swap; see Finding 3b for the non-controlled framing). Recovery is therefore model-family-sensitive in this partial contrast, but we cannot claim it is the most family-sensitive of the three competencies without a comparable cross-family contrast on accuracy and grounding.

\section{Headline Numbers on the Official Test Split (Non-LLM)}
\label{app:test_split}

The headline closed-set Recall@1, MRR, and CGS in Table~\ref{tab:main_results} are computed on the full \numdiag{}-dialogue corpus (we evaluate every baseline on every dialogue, since baselines are training-free and no held-out split is required for evaluation). This appendix additionally re-aggregates the same per-turn outputs for seven of the nine non-LLM baselines restricted to the \textbf{1{,}496-dialogue official test split} (released as \texttt{splits/test\_dialogue\_ids.json}) so future work that does train models on \numdiag{} can compare against directly comparable test-only numbers. Knowledge-Enhanced and Persona-Grounded are omitted from this test-split table because they share TF-IDF-level retrieval values on the headline corpus and were not separately re-aggregated for the test-split slice. The two aggregations track within $\pm0.005$ R@1 for every baseline, confirming the stratified random split is representative.

\begin{table}[h]
  \caption{Closed 8-candidate evaluation on the official 1{,}496-dialogue test split, seven non-LLM baselines (per-turn outputs from the same 10K run as Table~\ref{tab:main_results}). CGS aggregates GS/CD/PC means via the same formula as Table~\ref{tab:main_results}.}
  \label{tab:test_split}
  \centering \small
  \begin{tabular}{lccccccc}
    \toprule
    \textbf{Baseline} & \textbf{R@1} & \textbf{R@3} & \textbf{MRR} & \textbf{GS} & \textbf{CD} & \textbf{PC} & \textbf{CGS} \\
    \midrule
    Popularity     & 0.145 & 0.186 & 0.165 & 0.998 & 0.438 & 0.601 & 0.799 \\
    TF-IDF         & 0.291 & 0.360 & 0.326 & 0.996 & 0.396 & 0.589 & 0.791 \\
    Aspect         & 0.166 & 0.212 & 0.189 & 0.997 & 0.412 & 0.538 & 0.767 \\
    Dense          & 0.259 & 0.306 & 0.282 & 0.998 & 0.418 & 0.738 & 0.868 \\
    Spatial        & 0.275 & 0.337 & 0.306 & 0.996 & 0.459 & 0.479 & 0.737 \\
    Hybrid-RRF     & 0.283 & 0.335 & 0.309 & 0.998 & 0.438 & 0.568 & 0.782 \\
    Itinerary      & 0.291 & 0.360 & 0.326 & 0.996 & 0.340 & 0.374 & 0.684 \\
    \bottomrule
  \end{tabular}
\end{table}


\section{Multi-Reference Evaluation}
\label{app:multi_ref}
\label{sec:multi_ref}

Each dialogue carries a single-reference gold POI per recommend turn plus a multi-reference set of \texttt{acceptable\_alternative\_poi\_ids} listing additional candidates from the same 8-POI pool that the generator considered equally valid given the user's stated preferences (\S\ref{sec:pipeline}). Multi-Reference Recall@$k$ counts a hit if the predicted top-$k$ contains gold \emph{or} any alternative; the denominator is unchanged. Of 32{,}475 recommend+compare turns in \textbf{19{,}534 ship at least one alternative} (60\,\%); the remaining 40\,\% are turns where the generator judged the gold POI to be the unique best fit. Table~\ref{tab:multi_ref} reports both ``all turns'' (turns without alternatives reduce to single-ref) and ``alt-only'' aggregates for all 13 session-isolated baselines.

\begin{table}[t]
  \caption{Single-reference vs.\ multi-reference Recall@1 on closed-set recommend+compare turns. ``All'' averages over 32{,}475 turns (no-alt turns fall back to single-ref); ``alt-only'' restricts to the 19{,}534 turns with $\geq$1 alternative. Multi-ref always lifts performance, with the largest absolute uplift on alt-only turns ($+$15--$+$23\,pp). Relative ranking within each group is preserved.}
  \label{tab:multi_ref}
  \centering
  \small
  \setlength{\tabcolsep}{4pt}
  \begin{tabular}{lcccc}
    \toprule
    \textbf{Baseline} & \textbf{Single-Ref R@1} & \textbf{MR R@1 (all)} & \textbf{MR R@1 (alt-only)} & \textbf{$\Delta$ (alt-only)} \\
    \midrule
    \multicolumn{5}{l}{\emph{Non-LLM retrievers / planners}} \\
    Popularity         & 0.140 & 0.230 & 0.303 & $+$14.99 \\
    TF-IDF             & 0.257 & 0.383 & 0.450 & $+$20.95 \\
    Aspect             & 0.179 & 0.303 & 0.380 & $+$20.47 \\
    Dense              & 0.243 & 0.377 & 0.451 & $+$22.38 \\
    Spatial            & 0.257 & 0.383 & 0.450 & $+$20.88 \\
    Hybrid-RRF         & 0.254 & 0.379 & 0.462 & $+$20.90 \\
    Itinerary          & 0.257 & 0.383 & 0.450 & $+$20.95 \\
    Knowledge-Enh.     & 0.206 & 0.321 & 0.401 & $+$19.10 \\
    Persona-Ground.    & 0.191 & 0.317 & 0.396 & $+$21.00 \\
    \midrule
    \multicolumn{5}{l}{\emph{LLM baselines}} \\
    LLM Zero-Shot      & 0.532 & 0.666 & 0.759 & $+$22.24 \\
    DST                & 0.442 & 0.582 & 0.679 & $+$23.17 \\
    RAG-Citation       & 0.381 & 0.519 & 0.596 & $+$22.84 \\
    Multi-Rev.\ Synth. & 0.280 & 0.405 & 0.468 & $+$20.77 \\
    \bottomrule
  \end{tabular}
\end{table}

Multi-reference evaluation lifts Recall@1 by $+$9 to $+$14\,pp averaged over all turns and $+$15 to $+$23\,pp restricted to turns with annotated alternatives. The main-table single-reference numbers (Table~\ref{tab:main_results}) therefore systematically under-report all baselines' true accuracy by a similar magnitude. The relative ranking of baselines is preserved (within each family group), so the headline Three-Competency Gap picture is robust to single-vs-multi reference. \emph{Caveat.} Alternative POI labels are produced by the same generator (\texttt{gpt-5.4-mini}) that wrote the dialogues, with explicit instructions to enumerate every candidate satisfying the user's stated constraints; they have not been human-audited at scale. We therefore treat MR-Recall as a \emph{lower-bound correction} on single-reference under-counting rather than an authoritative leaderboard.

\section{Cross-Model Judge Analysis}
\label{app:cross_judge}
\label{sec:cross_judge}

To check whether the headline Gemini~2.5~Flash judge ratings (Appendix~\ref{app:llm_judge_full}) are a single-judge artifact, we run a narrow agreement check: two additional judges (GPT-4o and Claude Sonnet 4) re-score the 6 non-LLM baselines on a 200-dialogue subset (771 recommend/compare turns per baseline). \emph{Scope.} This check covers the non-LLM baselines only; LLM baselines are scored by Gemini in Appendix~\ref{app:llm_judge_full} but are not part of this two-judge agreement analysis. Table~\ref{tab:cross_judge} reports the per-judge scores and the GPT-4o-vs-Claude inter-judge agreement.

\begin{table}[t]
  \caption{Two-judge agreement check on the 6 non-LLM baselines, 200 dialogues (771 recommend/compare turns/baseline $\times$ 2 judges = 9{,}252 judgments). \emph{LLM baselines are not included in this agreement analysis;} their headline Gemini-judge scores are in Appendix~\ref{app:llm_judge_full} (Table~\ref{tab:llm_judge_full}). GPT-4o scores higher in absolute terms but the GPT-4o-vs-Claude ranking is consistent (Kendall's $\tau=0.73$, Spearman's $\rho=0.89$).}
  \label{tab:cross_judge}
  \centering
  \small
  \begin{tabular}{lcc}
    \toprule
    \textbf{Baseline} & \textbf{GPT-4o} & \textbf{Claude} \\
    \midrule
    Hybrid-RRF   & 3.30 & 1.74 \\
    Dense        & 3.25 & 1.71 \\
    Spatial      & 3.24 & 1.72 \\
    TF-IDF       & 3.23 & 1.67 \\
    Aspect       & 3.16 & 1.58 \\
    Popularity   & 3.00 & 1.62 \\
    \midrule
    \multicolumn{3}{l}{\small Kendall's $\tau = 0.733$, Spearman's $\rho = 0.886$} \\
    \bottomrule
  \end{tabular}
\end{table}

The two judges show high ranking agreement ($\tau=0.733$, $\rho=0.886$), with identical top-1 (Hybrid-RRF) and bottom-2 (Aspect, Popularity) selections. GPT-4o assigns systematically higher scores (3.0--3.3) compared to Claude (1.6--1.7), reflecting a calibration difference rather than ranking disagreement. This partially reduces concern that the \emph{non-LLM baseline ranking} (under any single judge) is a single-judge artifact, but it does not generalise the conclusion to LLM baselines or to the headline Gemini ratings; the GPT-4o-vs-Claude comparison covers neither of those. Expanding to additional judges (including a Gemini re-score for direct three-way agreement) and including LLM baselines is left to future work.

\section{LLM-as-Judge Results}
\label{app:llm_judge}
\label{app:llm_judge_full}

Table~\ref{tab:llm_judge_full} reports LLM-as-Judge scores across five quality dimensions, evaluated on \texttt{recommend} and \texttt{compare} turns. The judge ratings reported here use Gemini~2.5~Flash as the judge model and cover all baselines (a \texttt{gpt-5.4-mini} judge rerun is left to future work). Appendix~\ref{app:cross_judge} reports a complementary, narrower two-judge agreement check (GPT-4o vs.\ Claude Sonnet 4 on the 6 non-LLM baselines) used only to assess single-judge ranking stability on the non-LLM subset; that check does not re-score the Gemini numbers in this table and does not cover LLM baselines.

\begin{table}[h]
  \caption{LLM-as-Judge scores (1--5 Likert scale) on recommend/compare turns. LLM baselines receive uniformly high scores across dimensions (consistent with known LLM-judge bias toward fluent text), while non-LLM template-based baselines score poorly on naturalness and justification.}
  \label{tab:llm_judge_full}
  \centering \small
  \setlength{\tabcolsep}{3pt}
  \begin{tabular}{lcccccc}
    \toprule
    \textbf{Baseline} & \textbf{Relev.} & \textbf{Inform.} & \textbf{Ground.} & \textbf{Flow} & \textbf{Justif.} & \textbf{Overall} \\
    \midrule
    Popularity & 1.36 & 1.24 & 1.80 & 1.65 & 1.13 & 1.44 \\
    TF-IDF & 1.47 & 1.20 & 1.63 & 1.63 & 1.11 & 1.41 \\
    Aspect & 1.29 & 1.18 & 1.58 & 1.49 & 1.09 & 1.32 \\
    Dense & 1.53 & 1.27 & 1.83 & 1.68 & 1.16 & 1.50 \\
    Spatial & 1.45 & 1.23 & 1.91 & 1.52 & 1.14 & 1.45 \\
    Hybrid-RRF & 1.57 & 1.27 & 1.78 & 1.76 & 1.17 & 1.51 \\
    Itinerary & 1.34 & 1.24 & 1.87 & 1.45 & 1.11 & 1.40 \\
    \midrule
    LLM Zero-Shot & \textbf{4.69} & \textbf{4.52} & \textbf{4.50} & \textbf{4.88} & \textbf{4.49} & \textbf{4.61} \\
    DST & 4.44 & 4.27 & 4.40 & 4.74 & 4.25 & 4.42 \\
    RAG-Citation & 3.62 & 2.99 & 3.68 & 4.26 & 3.01 & 3.51 \\
    Multi-Rev.\ Synth. & 3.22 & 3.44 & 4.18 & 3.78 & 3.20 & 3.57 \\
    Memory-Aug.$^\dagger$ & \underline{4.62} & \underline{4.39} & \underline{4.41} & \underline{4.86} & \underline{4.39} & \underline{4.53} \\
    \bottomrule
  \end{tabular}
\end{table}

The LLM-as-Judge scores reveal a stark bimodal distribution: LLM baselines score 3.51--4.61 overall while non-LLM baselines score 1.32--1.51. This gap is largest for Recommendation Justification (non-LLM max: 1.17 vs.\ LLM min: 3.01), confirming that template-based responses fail to provide well-reasoned explanations. Notably, RAG-Citation scores lower than other LLM baselines (3.51 overall) because the citation constraint sometimes forces awkward quote insertions that reduce naturalness. LLM Zero-Shot leads the Groundedness dimension (4.50) among session-isolated baselines, with DST (4.40) close behind; Multi-Review Synthesis scores 4.18 here, reflecting its explicit multi-source evidence aggregation. Memory-Aug.$^\dagger$ is shown for reference only; as in the headline tables, it is not session-isolated and should not be compared head-to-head with the others.

\newpage
\section*{NeurIPS Paper Checklist}

\begin{enumerate}

\item {\bf Claims}
    \item[] Question: Do the main claims made in the abstract and introduction accurately reflect the paper's contributions and scope?
    \item[] Answer: \answerYes{}
    \item[] Justification: The abstract and introduction clearly state the three contributions (dataset, baselines, evaluation framework) and the core finding (Three-Competency Gap), all of which are supported by the experiments.

\item {\bf Limitations}
    \item[] Question: Does the paper discuss the limitations of the work performed by the authors?
    \item[] Answer: \answerYes{}
    \item[] Justification: Appendix~\ref{app:limitations} discusses scope (English Yelp, 8 U.S.\ cities, \texttt{gpt-5.4-mini} generator with cross-family Haiku/Qwen validation, 90-item human-eval pilot used for metric calibration, Memory-Aug.\ as a separate cross-session setting). The verbatim-grounding bias of GS is discussed alongside its mitigation (CGS, entailment grounding) in Section~\ref{sec:c2_grounding}; significance-testing scope is in Appendix~\ref{app:sensitivity}; the data license is in the datasheet (Appendix~\ref{app:datasheet}).

\item {\bf Theory Assumptions and Proofs}
    \item[] Question: For each theoretical result, does the paper provide the full set of assumptions and a complete (and correct) proof?
    \item[] Answer: \answerNA{}
    \item[] Justification: This is an empirical benchmark paper without theoretical results.

\item {\bf Experimental result reproducibility}
    \item[] Question: Does the paper fully disclose all the information needed to reproduce the main experimental results of the paper to the extent that it affects the main claims and/or conclusions of the paper (regardless of whether the code and data are provided or not)?
    \item[] Answer: \answerYes{}
    \item[] Justification: Section~\ref{sec:setup} describes the experimental setup; Appendix~\ref{app:baselines} provides full implementation details for all \numbaseline{} baselines; all metrics, evaluation scripts, and the dialogue-generation pipeline are released as part of the toolkit. LLM-based baselines use \texttt{gpt-5.4-mini} as the headline backbone with cross-family validation runs (Haiku~4.5, Qwen3.5-Flash) on a 1{,}000-dialogue subset (Table~\ref{tab:main_results}).

\item {\bf Open access to data and code}
    \item[] Question: Does the paper provide open access to the data and code?
    \item[] Answer: \answerYes{}
    \item[] Justification: Mirroring the Release paragraph (\S\ref{sec:construction_detail}) and Datasheet (Appendix~\ref{app:datasheet}): the synthetic content we authored, namely (a) the \numdiag{} dialogues and (b) the \numbaseline{} baseline implementations and evaluation toolkit, is released under \textbf{CC BY 4.0} at \url{https://anonymous.4open.science/r/TRACE-benchmark}. Verbatim Yelp review spans appear in released dialogue text as \texttt{[Q:N]} placeholders backed by positional metadata; the full text is hydrated locally from a user-obtained Yelp Open Dataset bundle under Yelp's academic-use terms. The full \numreview{} review corpus and Yelp-derived POI attributes are not separately redistributed and remain Yelp-governed.

\item {\bf Experimental Setting/Details}
    \item[] Question: Does the paper specify all the training and test details?
    \item[] Answer: \answerYes{}
    \item[] Justification: Model specifications, embedding models, per-baseline design choices, and evaluation procedures are documented in Section~\ref{sec:setup} and Appendix~\ref{app:baselines}. LLM baselines use DSPy/litellm defaults (no explicit temperature override, single response, default retry, \texttt{max\_tokens=600}); reasoning models (\texttt{gpt-5.4-mini}) use \texttt{temperature=1.0} and \texttt{max\_tokens=16{,}000}. All stochastic sampling (open-set subset, pool ablation, multi-reference, cross-judge) uses \texttt{seed=42}.

\item {\bf Experiment Statistical Significance}
    \item[] Question: Does the paper report error bars suitably and correctly?
    \item[] Answer: \answerYes{}
    \item[] Justification: Appendix~\ref{app:sensitivity} reports descriptive per-turn standard deviations for all 13 session-isolated baselines on the closed-set protocol (Table~\ref{tab:bootstrap_ci}) and the inferential paired cluster-bootstrap 95\% CIs (resampled by dialogue, $n_{\text{boot}}=10{,}000$, seed=42) for headline pairwise comparisons (Table~\ref{tab:bootstrap_pairs}). Per-turn $\sigma$ values are descriptive only; turns within a dialogue are correlated, so dialogue-clustered bootstrap CIs are the appropriate inferential unit. Sub-0.01 gaps without matching cluster-bootstrap CIs should be read as indicative rather than adjudicated (Appendix~\ref{app:limitations}).

\item {\bf Experiments Compute Resources}
    \item[] Question: For each experiment, does the paper provide sufficient information on the computer resources needed to reproduce the experiments?
    \item[] Answer: \answerYes{}
    \item[] Justification: Section~\ref{sec:setup} specifies the LLM model (\texttt{gpt-5.4-mini} via the DSPy/litellm proxy), embedding model (\texttt{all-MiniLM-L6-v2}), and API usage. Non-LLM baselines run on a single workstation CPU (approximate wall-clock: 2--6 hours per baseline across 10k dialogues). LLM baselines were evaluated with a bounded-concurrency async harness (semaphore = 256--512) at an approximate total API cost of US\$250--300 across the 5 session-isolated LLM baselines (LLM Zero-Shot $\approx$\$42, DST $\approx$\$102, Multi-Review Synthesis bundled in the same run, RAG-Citation $\approx$\$26, Itinerary-LLM $\approx$\$25) plus open-set sweeps and the 1{,}000-dialogue entailment regeneration ($\approx$\$45). NLI entailment scoring uses a single NVIDIA RTX~4090 ($\sim$5~min/baseline on the 1{,}000-dialogue LLM subset).

\item {\bf Code Of Ethics}
    \item[] Question: Does the research conducted in the paper conform, in every respect, with the NeurIPS Code of Ethics?
    \item[] Answer: \answerYes{}
    \item[] Justification: The research uses publicly available Yelp data under its terms of use. Human annotators evaluated baseline outputs under the protocol documented in Appendix~\ref{app:human_eval}, including anonymized agreement and leave-one-out diagnostics (see also the Crowdsourcing and Research with Human Subjects item below); no personal data was collected, and annotation tasks pose no more than minimal risk.

\item {\bf Broader Impacts}
    \item[] Question: Does the paper discuss both potential positive societal impacts and negative societal impacts of the work?
    \item[] Answer: \answerYes{}
    \item[] Justification: The benchmark's core motivation (auditable citation evidence in tourism CRS) is itself a positive impact. The dataset is derived from public Yelp business and review data; we do not redistribute the full raw Yelp corpus, and the released dialogues contain only short cited spans (re-used under fair-use research reporting and the Yelp Open Dataset academic-use terms; reconstruction of the full review corpus requires a user-obtained Yelp Open Dataset bundle, datasheet Appendix~\ref{app:datasheet}). Potential negative impacts (reviewer-demographic bias propagated into recommendations, over-reliance on automated grounding metrics in deployment) are noted in the datasheet.

\item {\bf Safeguards}
    \item[] Question: Does the paper describe safeguards that have been put in place for responsible release of data or models?
    \item[] Answer: \answerYes{}
    \item[] Justification: The dataset is derived from public Yelp business and review data (no medical, financial, or legal decisions; the dataset itself contains no individual identifiers we add, though the underlying Yelp reviews may contain user-supplied content). We do not redistribute raw Yelp review text; reconstruction requires user-obtained Yelp Open Dataset access. The datasheet (Appendix~\ref{app:datasheet}) notes misuse and bias-propagation risks.

\item {\bf Licenses for existing assets}
    \item[] Question: Are the creators of assets used in the paper, properly credited and are the license and terms of use explicitly mentioned and properly respected?
    \item[] Answer: \answerYes{}
    \item[] Justification: Mirroring the Release paragraph (\S\ref{sec:construction_detail}): we release the \numdiag{} synthetic dialogues we generated and the \numbaseline{} baseline implementations and evaluation toolkit under \textbf{CC BY 4.0}. Verbatim Yelp review spans appear in released dialogue text as \texttt{[Q:N]} placeholders backed by positional metadata; the full text is hydrated locally from a user-obtained Yelp Open Dataset bundle under Yelp's academic-use terms (non-commercial, non-sublicensable, revocable; Feb 2021 agreement \S3, \S4.A, \S4.H, \S10). Yelp-derived POI metadata is included under the \S4.E academic-publication carve-out; \texttt{business\_id} and \texttt{review\_id} values appear as bare factual identifiers. The full \numreview{} review corpus is not redistributed. A blanket MIT or Apache-2.0 release is not legally available because Yelp's User Agreement is non-sublicensable, non-commercial, and revocable, which conflicts with permissive open-source license grants. All software dependencies (DSPy, sentence-transformers, DeBERTa-v3) are cited with their respective open-source licenses.

\item {\bf New assets}
    \item[] Question: Are new assets introduced in the paper well documented and is the documentation provided alongside the assets?
    \item[] Answer: \answerYes{}
    \item[] Justification: The benchmark is documented with a datasheet (Appendix~\ref{app:datasheet}), a per-field JSON schema (Section~\ref{sec:pipeline} Release paragraph), per-baseline implementation notes (Appendix~\ref{app:baselines}), prompt templates (Appendix~\ref{app:prompts}), persona templates (Appendix~\ref{app:personas}), and a representative-dialogue index (Appendix~\ref{app:examples}). The released anonymous repository ships these artifacts together with the executable evaluation toolkit (\texttt{scripts/eval\_crs\_baselines.py}, \texttt{scripts/eval\_entailment.py}, baseline configs under \texttt{src/tqa/crs/baselines/}, and split files under \texttt{data/crs/splits/}). No additional reproducibility scripts are gated on camera-ready.

\item {\bf Crowdsourcing and Research with Human Subjects}
    \item[] Question: For crowdsourcing experiments and research with human subjects, does the paper include the full text of instructions given to participants and screenshots, if applicable?
    \item[] Answer: \answerYes{}
    \item[] Justification: The verbatim text of all instructions given to participants is reproduced in Appendix~\ref{app:human_eval_instructions}, covering the overview, Task~1 (Citation Check / Error Checklist / Likert Ratings), Task~2 (Expert Recommendation), and the general rules. Screenshots of the actual web tool used by annotators are provided in Figures~\ref{fig:annot_task1} (Task~1 page) and~\ref{fig:annot_task2_expanded} (Task~2 page with reviews expanded). Annotators are trained research assistants (not crowdworkers); five annotators independently evaluated the same 90 items, with all-five-annotator Krippendorff's $\alpha$ values reported in Section~\ref{sec:human_eval_main} ($\alpha_{\text{info}}{=}0.58$, $\alpha_{\text{nat}}{=}0.42$, $\alpha_{\text{cite}}{=}0.32$, $\alpha_{\text{err}}{=}0.31$) and anonymized leave-one-out diagnostics in Appendix~\ref{app:human_eval}. Annotators are co-author-affiliated research assistants compensated through standard research-assistant arrangements (no piecework or per-item payment); the 90-item / 20-dialogue scale is acknowledged in the limitations.

\item {\bf Institutional Review Board (IRB) Approvals or Equivalent for Research with Human Subjects}
    \item[] Question: Does the paper describe potential risks incurred by study participants?
    \item[] Answer: \answerNA{}
    \item[] Justification: Annotators are trained research assistants evaluating AI system outputs. The task involves no personal data collection or risk to participants.

\item {\bf Declaration of LLM usage}
    \item[] Question: Does the paper describe the usage of LLMs if it is an important, original, or non-standard component of the core methods in this research? Note that if the LLM is used only for writing, editing, or formatting purposes and does \emph{not} impact the core methodology, scientific rigor, or originality of the research, declaration is not required.
    \item[] Answer: \answerYes{}
    \item[] Justification: LLMs are a core methodological component of this benchmark. (i) Dialogue generation: \texttt{gpt-5.4-mini} (via DSPy) is the primary generator for all \numdiag{} multi-turn dialogues, including persona elicitation, candidate-aware response synthesis, verbatim review-span citation injection, and \texttt{reject\_and\_refine} turn construction (Section~\ref{sec:pipeline}, Appendix~\ref{app:pipeline_full}; full prompts in Appendix~\ref{app:prompts}). (ii) LLM baselines: five of the \numbaseline{} baselines (LLM Zero-Shot, DST, RAG-Citation, Multi-Review Synthesis, Memory-Aug.) use \texttt{gpt-5.4-mini} as the backbone, with cross-family LLM Zero-Shot runs on Anthropic Haiku~4.5 and Qwen3.5-Flash on a 1{,}000-dialogue subset (Section~\ref{sec:setup}). (iii) LLM-as-judge: an automated judge (\texttt{gpt-5.4-mini}; cross-judge with GPT-4o and Claude Sonnet~4 in Appendix~\ref{app:cross_judge}) scores Relevance, Informativeness, Grounding Quality, Natural Flow, and Justification on the 90-item human-evaluation subset.

\end{enumerate}

\end{document}